\documentclass[11pt]{article}   
\usepackage{amsthm,amsmath,amssymb,mathrsfs,cite,amscd}
\numberwithin{equation}{section}

\def\to{\mathchoice
{\longrightarrow}
{\rightarrow}
{\rightarrow}
{\rightarrow}}

\newcommand{\p}{\partial}
\newcommand{\<}{\langle}
\renewcommand{\>}{\rangle}

\newcommand{\half}{\tfrac12}

\newcommand{\CF}{\mathcal{F}}
\newcommand{\CH}{\mathcal{H}}

\newcommand{\CK}{\mathcal{K}}

\newcommand{\CM}{\mathcal{M}}
\newcommand{\CN}{\mathcal{N}}
\newcommand{\CO}{\mathcal{O}}

\newcommand{\CZ}{\mathcal{Z}}

\newcommand{\virt}{{\textup{virt}}}

\renewcommand{\geq}{\geqslant}
\renewcommand{\leq}{\leqslant}
\newcommand{\C}{\mathbb{C}}
\newcommand{\E}{\mathbb{E}}

\newcommand{\N}{\mathbb{N}}
\renewcommand{\P}{\mathbb{P}}
\newcommand{\Q}{\mathbb{Q}}
\newcommand{\R}{\mathbb{R}}
\newcommand{\Z}{\mathbb{Z}}

\newcommand{\Li}{\textrm{Li}}

\renewcommand{\o}{\otimes}
\renewcommand{\half}{\frac{1}{2}}
\newcommand{\tr}{{\rm tr}}
\newcommand{\str}{{\rm str}}

\renewcommand{\th}{\vartheta}
\newcommand{\mbar}{\overline{\mathcal{M}}} 
\newcommand{\bs}{\begin{split}}
\newcommand{\be}{\begin{equation}}
\newcommand{\es}{\end{split}} 
\newcommand{\ee}{\end{equation}}

\theoremstyle{plain}

\theoremstyle{remark}

\newtheorem{ex}[equation]{Example}

\begin{document}

\begin{titlepage}   \thispagestyle{empty}
  \title{\Large \bf Gromov-Witten Theory and Threshold Corrections}
   \author{Daniel B. Gr\"unberg \\ {\em \small KdV Institute, Plantage
   Muidergracht 24, 1018 TV Amsterdam, The Netherlands} \\ \small
   grunberg@mccme.ru}  
   \date{July 2004}
   \maketitle
   \vspace{2cm}
   \begin{abstract}  \large
     We present an overview of Gromov-Witten theory and its links with
     string theory compactifications, focussing on the GW potential as
     the generating function for topological string amplitudes at
     genus $g$.  Restricting to Calabi-Yau target spaces, we give a
     complete derivation of the GW potential, discuss problems of
     multicovers and the infinite product expression.  We explain the
     link with counting instantons or BPS states in type IIA and
     heterotic string theories.  We show why the numbers of BPS states
     on the heterotic side can be a priori expressed in terms of those
     on the type IIA side, and vice versa. We compute heterotic
     one-loop integrals to obtain the genus $g$ GW potential, and
     detail two ways to obtain threshold corrections for heterotic
     orbifolds, a prerequisite for the notorious work by Harvey and
     Moore.  We review this long and cumbersome construction in a
     self-contained way and make it explicit in examples of
     compactifications.  We also develop the relation to Jacobi forms
     and automorphic forms, and clarify the meaning of the
     Gopakumar-Vafa invariants.
  \end{abstract}
\end{titlepage}

\small
\tableofcontents      
\thispagestyle{empty}  
\normalsize
\newpage

\section*{Introduction} \label{sec:intro}
\addcontentsline{toc}{section}{Introduction}

Gromov-Witten (GW) theory is usually considered to have its orgin in
Witten's seminal work \cite{W-91} on $2d$ gravity where he solved
integrals of algebraic geometry having the enumerative meanings of
counting instantons (or holomorphic curves embedded in a target
space).  The thirteen-year-old theory has not merely reached the age
of reason, it has also escalated into an independent and explosive
field of study.  What was originally a string theoretic problem and
approach, has now almost been hijacked by algebraic geometers, subdued
to rigour and held to leading strings.  And gladly.  For the theory
now stands on a firm rock, albeit an arduous one to clamber for the.
This makes it difficult for physisists nowadays to understand what was
once a topic of their own, of instanton sums in topological field
theory.

\subsubsection*{GW Theory}
\addcontentsline{toc}{subsubsection}{GW Theory} 

In GW theory, one studies holomorphic maps $f: \Sigma_g \to X $ from a
Riemann surface $ \Sigma_g$ to a target space $X$, also known as
instantons.  These give non-perturbative contributions to the action
of an $N=2$ sigma-model defined on $X$ whose twisting yields a
topological field theory (A-model).  The latter's amplitudes $F_g (t)$
at genus $g$ are called topological string partition functions in the
sense that they do not depend on the worldsheet ($\Sigma_g$) metric
but only on the cohomology class of the spacetime metric (K\"ahler
moduli $t$ of the target).

The image of the map will be a holomorphic curve with a homology
class, singular points, an arithmetic genus, a geometric genus, etc.
The aim is to ``count'' such objects in $X$; it turns out it is easier
to ``count maps'' with domain a well-behaved Riemann surface (that
possibly degenerates).  Roughly speaking, the ``number of maps''
corresponds to the GW invariant (a rational number), while the
``number of (image) curves'' is our instanton number (an integer). 

Of course, these maps or curves come typically in infinite families,
hence the need to impose constraints such as: the curve should pass
through a given hypersurface $Z$ of $X$, and its cohomological dual
$Z^* =\gamma$ is a 2-form.  An appropriate array of constraints thus
yields hopefully a finite number, though a rational one, as the moduli
spaces we work with are orbifolds and there will be a trailing
denominator from the order of some automorphism group.  From these
rational numbers one can extract {\em integer numbers} by a recursion
involving {\em excess intersection}; these are our sought-for
instanton numbers. 

For {\em closed} string instantons, and we shall only consider such,
the physical approach for counting them is paralleled by a
well-defined theory developed by algebraic geometers.  The latter use
moduli spaces of stable maps, on which they integrate pull-backs of
differential forms $\gamma_i$ on the target space, to define the {\em
  GW invariants}.  {\em Calabi-Yau threefolds} are singled out in
nature as not requiring any constraints (to obtain a finite number of
maps) and hence requiring no $\gamma_i$ either.  Their GW invariants
thus only depend on the genus $g$ of the domain curve and the homology
class $\beta$ (or $d$) of the image curve; we denote them by $N^g_d$
or $\< 1 \>_{g,\beta}$.

These rational numbers are linearly related to the {\em instanton
  numbers} (or {\em BPS invariants}) $n^r_d$.  That the latter are
{\em integers} is a non-trivial fact met in all examples known so far,
shadowy for mathematicians but natural for physicists who see in them
numbers of D-branes (or BPS states) wrapped around particular cycles
of the CY threefold.

\subsubsection*{GW Potential for Type IIA and Heterotic Strings}
\addcontentsline{toc}{subsubsection}{GW Potential for Type IIA and
  Heterotic Strings} 

The GW invariants at genus $g$ can be gathered into {\em GW
  potentials} $F_g$. This is nothing but the genus $g$ amplitude of
the afore-mentioned topological strings.  The closest string theory
among the five $10d$ theories to describe this model is {\em type
  IIA}, and in this context the $F_g$ play the role of {\em couplings}
to the graviphoton field strength in the action.  They are exact at
genus $g$, and can thus be computed at strong coupling, also known as
the decompactification limit or ``M-theory'' (11-dimensional).

In the context of M-theory, a one-loop Schwinger integral will yield
the exact result for the $F_g$ (though $g$-loop was required in type
IIA), much as Seiberg-Witten theory is exact at one-loop.  Because of
strong coupling, only the lightest states will contribute to the
integral: D0- and D2 branes of type IIA, corresponding to KK-modes and
M2-branes of M-theory.  The {\em full} GW potential for the example of
the local conifold can then be re-written in product-notation and
yields the generating function for plane partitions.  Generalisations
of this example, including spin, can equally be presented in the form
of infinite products and it is hoped that these have {\em automorphic
  properties} of some sort.  The powers occurring in these products
contain our sought-for instanton numbers and go here by the name of
Gopakumar-Vafa invariants.

Another context where the $F_g$ are exact at one-loop is {\em
  heterotic string theory}, which is related to type IIA by {\em
  duality}.  The topological amplitudes also have the meaning of
couplings to the graviphoton field strength in the action, yet the
one-loop integral can be computed via the trick of lattice reduction,
yielding awkward expressions but whose holomorphic limit recovers the
desired constant term of the instanton generating function (this term
arising from constant maps is related to the {\em global} topology of
the space and appears thus with the Euler character).
  
Other prowesses from calculations in heterotic theory involve the full
determination of the prepotential, or $F_0$.  It goes via an ODE for
the gauge coupling that contains integrals of a theta function and a
modular form over the fundamental domain $\CF = \CH / SL(2,\Z)$.  The
latter integrals are resolved by the trick of unfolding the fundamental
domain and yield an astounding expression: the logarithm of
automorphic products \`a-la Borcherds.  Reading off this result as if
we were in type IIA string theory, we can extract our instanton
numbers or GW invariants.  In other words, there is a salient
relationship between GW invariants and automorphic forms -- at least
for the genus-0 invariants counting {\em rational} curves in the
target space.  Whether this holds for higher-genus invariants remains
opaque. 

Yet the string duality relating type IIA and heterotic theories still
provides an original way for presenting enumerative topological
information as powers in infinite product expressions: on the type IIA
side we have linear combinations of Gopakumar-Vafa invariants, while
on the heterotic we have powers of Borcherds products (that originate
from coefficients of nearly-holomorphic modular forms) and that are
parametrised by ``positive'' roots of even self-dual Lorentzian
lattices.

\subsubsection*{Bella Vista around the GW Potential}
\addcontentsline{toc}{subsubsection}{Bella Vista around the GW Potential}

The structures and mathematical tools hidden behind this long chain of
dualities, weak- and strong coupling limits and SCFTs are quite
enticing.  For instance, Borcherds' construction of automorphic forms
goes via lifting of {\em Jacobi forms} of index 1 (which are roughly
isomorphic to modular forms), and if the latter have zero weight the
former admit a product expression.  Examples of Jacobi forms include
the {\em elliptic genus}, a physical-topological object defined by a
trace over the Hilbert space of states of an SCFT: $\Phi(q,y) =$
tr$_{_{R,R}} (-1)^F q^{L_0-{c \over 24}} y^{J_0}$.  It has an explicit
expression known for several compactification spaces (CY
1,2,3,4-folds).

For a non-linear sigma model on the complex surface $K3$ (CY
two-fold), the above traces (which can be generalised for all
combinations of R or NS in the left- and right-moving sectors) yield
several topological indices (elliptic genus, Dirac genera, partition
functions, N=4 characters) that can be written as sums over a finite
number of `orbits' of the N=4 SCA.  

The starting point for the one-loop integrals in the heterotic context
are {\em threshold corrections} to gauge couplings.  These can be
computed in the effective field theory and the integrand will be
essentially a trace over the internal degrees of freedom.  At this
stage there is a road junction for the special example of heterotic
compactification on $K3 \times T^2$: we can either directly compute the
partition function of the model and find the trace over the internal
theory to yield $\Gamma_{10,2} E_6/\eta^{24}$, or we can rewrite the
abstract integrand using heavy algebra and landing on the so-called
``new susy index'' (a variant of the elliptic genus, closely related
to {\em helicity supertraces}).  From here, we can anew compute the
explicit value of the integrand using techniques from helicity
supertraces over BPS states (which tell us in particular that vector-
and hypermultiplets contribute equal and opposite amounts to the
supertrace), and we arrive at the same result of $\Gamma_{10,2}
E_6/\eta^{24}$.  This is yet without the additional factor from the
gauge group, a term involving quadratic Casimir operators and beta
functions.

In other words, the path towards writing down a one-loop threshold
correction, solving it and presenting it in a shape suitable to obtain
Gromov-Witten invariants is long and cumbersome, requires patience
from the reader as well as familiarity with many topics of string
theory and geometry.  It is the fondest hope of the author that this
long path is brought forward in the present article in a self-reliant
and comprehensible manner.

\subsubsection*{Acknowledgements}
\addcontentsline{toc}{subsubsection}{Acknowledgements}

This paper was written two years ago; I have not found time to complete
it and never will.
It is a pleasure to remember the many people who have helped me
with their valuable comments.  I acknowledge a myriad of discussions
that have contributed to my understanding of the subject and of
related topics during the past years, and shall only mention a few
names: J.~Bryan, R.~Pandharipande, M.~Ro\v cek and S.~Katz from the
PCMI school on Mirror Symmetry; J.~de Boer, A.~Klemm and B.~Pioline on
different occasions; and especially R.~Dijkgraaf.

I thank prof.~L\"ust for his hospitality at the Humboldt University in
Berlin (support from a `Kurzfristiges Forschungsstipendium' of the
DAAD), as well as prof.~Sossinsky for his hospitality at the
Independent Unversity of Moscow (support from the Netherlands
Organisation for Scientific Research (NWO)), and most of all the Racah
Institue of the Hebrew University whose financial support enabled me
to work on this project within Israel's vibrant string theory
community.

\section{Gromov-Witten Results on CY Threefolds} \label{sec:gw-results-CY}

This section restricts to Calabi-Yau threefolds (K\"ahler manifolds
with vanishing first Chern class).  The first two subsections deal
with the general theory (also valid for non-CY varieties) of GW
invariants; we recommend \cite{P-00} and \cite{CK-99} for further
details.  From the third subsection on, the formulae only make sense
for CY threefolds: we build the potentials $F_g$ and $F=\sum_g F_g
\lambda^{2g-2}$ based on the previous abstract definition of GW
invariants, but in fact most of it can be understood independently.
Though initially rather mathematical, this section later introduces
physical concepts such as the potential, the Yukawa couplings, etc,
all relating to the framework of {\em closed} string theory.

\subsection{The moduli space of stable maps}

Gromov-Witten theory studies the properties of curves embedded in
larger spaces via a holomorphic map $f:\Sigma_{g,n} \to X$.  We start
by introducing Kontsevich's powerful moduli space of stable maps, an
extension of the idea of moduli space $\mbar_{g,n}$ of stable nodal
curves $\Sigma_{g,n}$ of genus $g$ with $n$ markings $p_1,\cdots,
p_n$.  Recall that dim~$\mbar_{g,n} = 3g-3 +n $ ($g>1$).

A {\bf stable map} $f: \Sigma_{g,n} \to X$ is a map from a pointed
nodal curve to a variety $X$ such that the contracted components ({\em
  i.e.}~where $f$ is constant) of genus 0 have 3 special points and
those of genus 1 have 1 special point. (This ensures there won't be
any infinitesimal automorphisms, ie the automorphism group should be at
most finite.)

The {\bf moduli space} of stable maps is
$$
\mbar_{g,n}(X,\beta) := \{ \textrm{ stable maps } f: \Sigma \to X
~|~ f_*([\Sigma]) = \beta\in H_2(X,\Z) ~\}/\sim.
$$
Its elements are denoted $(\Sigma, p_1,\dots,p_n,f)$, and this is
{\em isomorphic} to $(\Sigma', p_1',\dots,p_n',f')$ iff ~$\exists
\tau: \Sigma \to \Sigma'$ with $\tau(p_i)=p_i'$ and $f'\circ\tau=f$.

Note that $\beta=0$ means constant maps.  If $H_2(X,\Z)=\Z$ (this is
the case of a hypersurface $X$ in $\P^N$ -- by the Lefschetz hyperplane
theorem), then we shall rather use the integer $d$ to label the class
$\beta=d \cdot \ell$ (where $\ell$ is the generator of $H^2$).  Note
also that a stable map need not have a stable domain curve, except if
$\beta=0$ where stable maps are equivalent to stable curves.
\begin{ex}
$$
\begin{array}{rl}
\mbar_{g,n}(X,0) &= \mbar_{g,n} \times X. \\
\mbar_{g,n}({\rm point},0) &= \mbar_{g,n} \\
\mbar_{0,0}(X,0) &= \mbar_{0,0} \times X =
\emptyset \times X = \emptyset \\
\mbar_{0,0}(\P^1,1) &= {\rm point} 
\end{array}
$$ 
\end{ex}
Note that $\CM_{g,n}(X,\beta)$ includes degree $d$ coverings of the
line (always singular for $d>1$, as the RH formula implies branch
points).
\begin{ex}
  $ \underline{\mbar_{0,0}(\P^N,1)} = \mbar_{0,0}(\P^N,1) =
  G(\P^1,\P^N)$, the Grassmannian of lines in $\P^N$, of dimension
  $2N-2$.  Note that $g=0$ implies that the geometric genus of
  $\Sigma_{0,0}$ also vanishes, leaving only a tree of $\P^1$'s as a
  possible candidate (if the $\P^1$'s form a loop, the arithmetic
  genus $g$ will increase by 1).  Together with $n=0$ and $d=1$, this
  forbids contractions (which would be unstable).  Embedding a tree in
  $\P^N$ means the total degree is the sum of the degrees of each
  component, {\em i.e.}~the number of lines in the tree; thus we can
  have only 1 line.
\end{ex}

\begin{ex} \label{d-fold-covers}
  $\underline{\mbar_{0,0}(\P^1,d)}$, $d>1$, is the space of all
  $d$-fold covers of the line, and has dimension $2(d-1)$ (all branch
  points).  This agrees with the dimension of the space of
  parametrised rational curves\footnote{In general, a (parametrised)
    {\em rational curve} of degree $d$ in $\P^N$ is a tuple of homogeneous
    polynomials of degree $d$ in $x_0, x_1$: $(f_0(x_0, x_1),\dots,
    f_N(x_0, x_1) )$, where the $f_i$ have no common factors.  The
    curve will be degenerate if it lies in a hyperplane; this happens
    in particular when some $f_i$ is 0 (hyperplane is $\{ y_i=0 \}$),
    or when $f_i =f_j$ (hyperplane is $\{ y_i-y_j=0 \}$), or in more
    general situations like $(x_0^2, ~x_1^2, -x_0^2 -x_1^2)$
    (hyperplane is the line $\{ y_0+y_1+y_2=0 \}$ in the plane, and
    the map is a
    double cover branched over $ (1,0,-1)$ and $ (0,1,-1)$ ), etc.  \\
    Rational curves can be nodal (increasing their arithmetic genus),
    as is the case of $(x_0^3, ~x_0x_1^2, ~x_1(x_1+x_0)(x_1-x_0))$,
    also written $(1,~z^2, ~z(z+1)(z-1)) =:(1,x,y)$ or $y^2=x(x-1)^2$,
    which is the cubic with an ordinary double node in $(1,1,0)$, {\em
      i.e.}  an elliptic curve ($g=1$). \\ The dimension of this space
    of rational curves is \underline{$(N+1)(d+1)-4$} (number of
    coefficients in the polynomials), where we have subtracted 1 for
    the redundancy in $\P^N$ and 3 for the reparametrisations of
    $(x_0,x_1)$. This number is much smaller than the dimensionality
    of all degree $d$ curves in $\P^N$, which are parametrised by
    ${N+d \choose d} -1$ coefficients (corresponding to polynomials of
    degree $d$ in $N+1$ variables).  The two dimensions match for
    $N=2$ and $d=1,2$, {\em i.e.}~all lines and conics in the plane
    can be parametrised as rational curves; for plane cubics,
    quartics,..., only a minority of them are rational.  Conversely, a
    rational curve of degree $d$ in $\P^2$ will in general be a plane
    curve of degree $D\geq 2d-3$ (need to solve $dD+1$ equations for
    $d(d+3)/2$ projective unknowns).} of degree $d$ to $\P^1$.  The
  boundary of this moduli space consists of all maps from trees with
  at least two lines (hence a nodal curve); adding contractions, the
  domain tree can have more than $d$ lines and the configuration can
  be very wild.
\end{ex}

We now define the {\em evaluation map} 
$$
\begin{array}{rrcl}
{\rm ev}_i :& \mbar_{g,n}(X,\beta) &\to& X \\
&(\Sigma,p_1,\dots,p_n,f)  &\mapsto& f(p_i)
\end{array}
$$
which allows us to present the {\em universal curve}
\begin{equation}  \label{evaluation-map}
  \begin{array}{ccccc}
&&\mbar_{g,n+1}(X,\beta) &&\\
& \pi \swarrow &&  \searrow {\rm ev}_{n+1} &\\
\mbar_{g,n}(X,\beta)&&&& f(p_{n+1}) \in X\\
\end{array}
\end{equation}
with which we will pull back objects from $X$ to $ \mbar_{g,n}(X,\beta)
$ (a standard trick).

The space $ \mbar_{g,n}(X,\beta) $ is a stack (orbifold), usually
singular and almost always will its deformation theory be obstructed.
It can consist of several components of different dimension, so we an
only speak of an expected or {\em virtual dimension}, vdim, which is a
lower bound to the actual local dimension. If all dimensions agree,
the theory is {\em unobstructed} and vdim = dim.  This number vdim is
defined as the deformations $h^0(\Sigma, f^* N^{\Sigma/X})$ of the
image curve $f(\Sigma)$ and is easily computed via Riemann-Roch in
case of no obstruction \cite{KM-94}:
\begin{equation}
  {\rm vdim} ~\mbar_{g,n}(X,\beta) = \int_\beta c_1(X) + ({\rm dim}~X
  -3) (1-g) +n . 
\end{equation}

\begin{ex} \label{ex:beta=0}
  When $\beta=0$, vdim $=$ dim $X(1-g) +$ dim $\mbar_{g,n}$, which
  matches the actual dimension (dim $X +$ dim $\mbar_{g,n}$) only
  for $g=0$.  Thus constant maps are obstructed at $g>0$.
\end{ex}
\begin{ex}
  If $X$ is a \underline{CY threefold} ({\em i.e.}~$c_1(X)=0$), the paragon
  for string theory, then \\ vdim~$\mbar_{g,0}(X,\beta) = 0$.
\end{ex}
\begin{ex}
  For degree $d$ rational curves in $\P^N$, we have $\int_\beta
  c_1(X) = \int_{d [\ell]} (N+1)H = d(N+1)$, and so \underline{vdim
  $\mbar_{0,0}(\P^N,d)$} $= (N+1)(d+1) -4$. This agrees with the
  actual dimension ($N+1$ polynomials of degree $d$), see footnote in
  Example~\ref{d-fold-covers}.
\end{ex}
\begin{ex} \label{ex:vdim-hypersurface}
  For a \underline{hypersurface $X$} of degree $D$ in $\P^N$, we have
  by the adjunction formula: $c_1(X) = -K_X = -(K_{\P^N} +[X] )_X=
  (N+1-D) H|_X$, where $H$ is the hyperplane class of $\P^N$.  Hence
  vdim~$\mbar_{g,n}(X,d) = d(N+1-D) +(N-4)(1-g) +n$.
\end{ex}

Similar to vdim, there will be a {\em virtual fundamental class}, of
dimension vdim and coinciding with the usual fundamental class in case
the theory is unobstructed.  If $\mbar_{g,n}(X,\beta)$ is
non-singular, the virtual fundamental class will turn out to be the
Euler class (top Chern class) of the obstruction bundle\footnote{The
  {\em obstruction sheaf} is the pull-back of the tangent sheaf of $X$
  to $\mbar_{g,n}(X,\beta)$ via (\ref{evaluation-map}), that is $
  R^1\pi_* {\rm ev}_{n+1}^* ~T_X $, the fibre of which is $H^1(\pi^* {\rm
    ev}_{n+1}^* ~T_X)$.  If the moduli space is non-singular, it is
  locally free ({\em i.e.}~a vector bundle).} Ob:
\begin{equation} \label{vir-fund-class-ns}
[ \mbar_{g,n}(X,\beta) ]^{\rm vir} = e({\rm Ob}) \cap [
\mbar_{g,n}(X,\beta) ] .
\end{equation}

\subsection{Gromov-Witten Invariants}

\subsubsection{Definition and Axioms}

We proceed to define GW invariants for general target spaces $X$.
These are invariant under complex structure deformations of $X$.  The
(primary) {\bf Gromov-Witten invariants} are the following correlators
involving cohomology classes $\gamma_i \in H^*(X,\Q)$:
  \begin{equation}  \label{gw-invariants}
      \< \gamma_1 \dots \gamma_n \>_{g,\beta} 
= \int_{ [\mbar_{g,n}(X,\beta)]^{\rm vir} } {\rm ev}_1^*(\gamma_1) 
    \cup \dots \cup {\rm ev}_n^*(\gamma_n) ,
\end{equation}
with the dimension matching condition \underline{$ \sum$ deg $\gamma_i =2$ vdim
  $\mbar_{g,n}(X,\beta)$} (otherwise vanishing result), so that the
result is a mere number (rational since our forms $\gamma_i$ are
$\Q$-valued).

The {\em enumerative meaning} of this number is as follows.  The
integrand is the cohomology class represented by those maps
$(\Sigma_g, p_1,\dots,p_n,f)$ satisfying $f(p_i) \in Z_i ~~\forall i$,
where $Z_i$ is a subvariety (of $X$) in the homology class dual to
$\gamma_i$.  Then the above GW invariant ``counts'' the number of
curves (or rather maps) satisfying this requirement together with
$f([\Sigma])=\beta$.  Because these curves usually come in infinite
families, the final integral to compute will be an Euler numer of a
bundle parametrising these families, yielding a rational number.

We now list three properties of GW invariants \cite{KM-94}:

\underline{(1) Fundamental class axiom:} If one of the $\gamma_i$ (say
$\gamma_n$) is the fundamental class $[X]=1_X \in H^0(X,\Q)$, the GW
invariant vanishes:
$$
\< \gamma_1 \dots \gamma_{n-1} ~[X] ~\>_{g,\beta} = 0, 
$$
since it can be rewritten with the same integrand, but with $\gamma_n$
dropped and hence capped against $\mbar_{g,n-1}(X,\beta)$; since this
space has one dimension less, the cohomology class will not be of top
degree anymore and the integral is zero.  This holds when $n\geq1$
(or $n+2g\geq 4$ for $\beta=0$).

\underline{(2) Divisor axiom:} If $\gamma_n$ is a 2-form with Poincar\'e
dual $[Z_n]$ (divisor class), the candidate curves for stable maps
will have their point $p_n$ mapped on $Z_n$, and $\beta \cap [Z_n]$
will be a zero-chain with ``number of points'' $\int_\beta \gamma_n
~\in \Q$.  This separate requirement in itself offers $\int_\beta
\gamma_n$ ``choices'' of stable maps satisfying $f(p_i) \in Z_i
~~(i=1,\dots,n-1)$, so
$$
 \< \gamma_1 \dots \gamma_n \>_{g,\beta} =\left( \int_\beta
 \gamma_n \right) ~~\< \gamma_1 \dots \gamma_{n-1} \>_{g,\beta}.
$$
This holds when $n\geq1$ (or $n+2g\geq 4$ for $\beta=0$).

\underline{(3) Constant map axiom:} For $\beta=0, ~g=0$, the requirement
$f(p_i) \in Z_i$ translates to $f(\Sigma) \in Z_1 \cap\dots\cap Z_n$.
Since $ [Z_1] \cap\dots\cap [Z_n] =\int_X \gamma_1\cup
\dots\cup\gamma_n$ we need $\sum$ deg $\gamma_i = 2$ dim $X$.  This
agrees with $\sum$ deg $\gamma_i = 2$ vdim $\mbar_{0,n}(X,0)$ only if
$n=3$ (as $\mbar_{0,3}=$ point).  Hence $\< \gamma_1 \dots \gamma_n
\>_{0,0} = 0$ except
\begin{equation}   \label{const-map-axiom}
\< \gamma_1 \gamma_2 \gamma_3 \>_{0,0} = \int_{\mbar_{0,3} \times X}
 ~{\rm ev}^*(\gamma_1 \otimes \gamma_2 \otimes \gamma_3) 
= \int_X \gamma_1 \cup \gamma_2 \cup \gamma_3
\end{equation}
This does not hold for $g>0$, as it would be an obstructed case ({\em i.e.}
vdim $\neq$ dim), see also section~\ref{sec:const-maps}.

\subsubsection{Example: GW invariants of $\P^1$:}

This is an easy example.  Note that vdim $\mbar_{g,n}(\P^1,d)
=2d+2g-2+n$.  We only deal with two cohomology classes: the point
class $[pt]$ generating $H^2(\P^1,\Z)=\Z$ and the fundamental class
$[\P^1]$ generating $H^0(\P^1,\Z)=\Z$.

For \underline{$\beta=0$} ({\em i.e.}~$d=0$), we have to satisfy $\sum$ deg
$\gamma_i = 2(2g-2+n)$, hence $g=0,1$ only.  At $g=0$ we have (by the
constant map axiom) $n=3$ and $\< [pt][\P^1] [\P^1] \>_{0,0}
=\int_{\P^1} [pt] =1 $.  At $g=1$ any $n$ is allowed but all $\gamma_i=
[pt]$, so (by the divisor axiom) $\< [pt]^n \>_{1,0}= \< [pt] \>_{1,0}
=\int_{\mbar_{1,1}(\P^1,0) \cong \P^1} [pt] =1$.

For \underline{$\beta \neq 0$} ({\em i.e.}~$d>0$), we have to satisfy $\sum$
deg $\gamma_i = 2(2d+2g-2+n)$; if $\gamma_i= [\P^1]$, the GW invariant
vanishes by the fundamental class axiom.  Hence all $\gamma_i= [pt]$;
this can only hold if $2d+2g-2=0$, {\em i.e.}~$d=1,~ g=0$: $\< [pt]^n
\>_{0,1}= \< [pt] \>_{0,1} = \< 1 \>_{0,1} =
\int_{\mbar_{0,0}(\P^1,1)} 1 =1$ because $[\mbar_{0,0}(\P^1,1)]$ is a
0-cycle of degree 1 (only 1 point in that moduli space), or
alternatively because $ \< [pt] \>_{0,1} = \int_{\mbar_{0,1}(\P^1,1)
  \cong \P^1} [pt] =1$.

All other GW invariants vanish, in particular all those for $d>1$ or
$g>1$.  This is due to dimensional reasons and not because there are
no maps of such degrees from such curves.  In fact there are too many
of them (infinite families) and our constraints $f(p_i) \in Z_i$ (a
representative of the point class) are trivial -- hence void.  In
order to end up with a finite number of maps, one needs rather to
impose a stronger condition like specifying the branch points and
their ramification indices.  This is the {\em{} Hurwitz problem} and
far from being solved.

\subsubsection{Example: Rational Curves on the Quintic}
\label{sec:quintic}

Our next example concerns $g=0$ invariants on CY threefolds ({\em{}
  i.e.}~rational curves).  For simplicity, we shall confine ourselves
to the quintic hypersurface in $\P^4$ ($h^{1,1}=1$), but most of our
formulae are easily adapted to more general cases.

Note first that this example is unique in that vdim $\mbar_{g,n}
(X,\beta) =n$, allowing GW invariants with $n=0$ since $\sum$ deg
$\gamma_i = 0$ can be satisfied.  That is, the absence of cohomology
classes gives the GW invariant $\< 1 \>_{g,d}$ the meaning of directly
``counting'' maps or, hopefully, curves in the quintic (without
needing to pass through some points).  The GW invariants at $g=0$ are
very simple and heavily restricted by $\sum$ dim $\gamma_i = 2n$.

For \underline{$\beta=0$} they are $\< \gamma_1\gamma_2 \gamma_3
\>_{0,0} =\int_X \gamma_1 \cup \gamma_2 \cup \gamma_3$; for the
quintic with hyperplane class $H \in H^2(X,\Z)$, line class $
\ell={1\over 5} H^2 \in H^4(X,\Z)=\Z$ and point class $ [pt]={1\over
  5} H^3 \in H^6(X,\Z)=\Z$ this vanishes except for
$$
\< H H H \>_{0,0} =5, \qquad \< H ~\ell ~[X] \>_{0,0} = \< [pt] [X]
[X] \>_{0,0} = 1,  \qquad \textrm{or} \qquad
\< \gamma_1 \gamma_2 ~[X] \>_{0,0} ~\textrm{ for } \gamma_i\in
H^3(X,\Z) .
$$

For \underline{$\beta \neq 0$}, $\< \gamma_1 \dots \gamma_{n-1} [X]
\>_{0,d} =0$ by the fundamental class axiom.  The only other
possibility is that all $\gamma_i$ are 2-forms ($H$ for the quintic):
$\< H \dots H \>_{0,d} = d^n ~\< 1 \>_{0,d}$ by the divisor axiom.
Hence all GW invariants are simply 0 or multiples of $ \< 1 \>_{0,d}$.
(The same conclusion is true at $g>0$ and for general CY threefolds.)

That vdim $\mbar_{0,0}(X,\beta)$ is 0 suggests that there is a finite
number of rational curves at each degree $d$ (Clemens
conjecture)\footnote{Yet the actual dimension of
  $\mbar_{0,0}(X,\beta)$ is non-zero since for each such curve
  $f:\Sigma_0\to X$, we have whole families of $k$-fold covers $f\circ
  g$ with $g:\P^1 \to \P^1$ of degree $k$.}.  This number \underline{$
  N_d := \< 1 \>_{0,d} $} of rational maps of degree $d$ coincides
with the degree of the 0-cycle $ [\mbar_{0,0}(X,\beta)]^{\rm vir}$.
Brute force computation of the above Euler numbers leads to despair
(was met with increasing complexity till degree 3); luckily Mirror
Symmetry found a smart way of tackling all numbers in one stroke
\cite{CdGP-91}: $N_1=2875, ~N_2=4876875/8$,\dots

\paragraph{Multicovers and Instanton Numbers:}

Despite ``counting'' rational maps, these invariants are generally
$\in \Q$ due to the orbifold structure of $\mbar_{0,0}(X,\beta)$ and
of $\mbar_{0,0}(\P^4,d)$.  Moreover, they include contributions from
$k$-fold covers of curves of degree $d/k$. An excess intersection
calculation \cite{AM-91} determines this contribution to be $1/k^3$
for each suc cover.  Define now the {\em instanton numbers} $n_d$
inductively by
\begin{equation}\label{quintic-rational}
  N_d = \< 1 \>_{0,d} =: \sum_{k|d} {1\over k^3}~n_{d/k}.
\end{equation}

In our context, the $n_d$ seem to play the role of numbers of smooth
isolated rational curves of degree $d$ in the
quintic.\footnote{Indeed, for $d\leq 9$ (where the Clemens conjecture
  is known to hold), this can be verified; but for $d=10$, problems
  arise due to double-covers of (6-nodal) degree 5 curves, and
  $n_{10}$ is not quite the number of rational curves \cite{V-93}.}  A
marvellous fact holds since Mirror Symmetry was able to generate all
$N_d$ in a row: {\em the $n_d$ are all integers} !

It is those instanton numbers that we shall associate with the numbers
of BPS states or the degeneracies of $sl(2,\C)$ spin representations
of such states wrapping bound states of D0- and D2-branes (see eqns
(\ref{physical-n^r_d}), (\ref{F}) and section~\ref{sec:physical-meaning}).

We can even understand the ``excess'' factor of $k^{-3}$ in
(\ref{quintic-rational}).  Had we rather computed $ \< H^3 \>_{0,d}$,
we would have counted the number of degree $d$ maps with the added
constraint of three image points lying on three different hyperplanes.
Again, contribution from multiple covers would have written this as
$\sum_{k|d}$ (number of degree $k$ curves)$\times$ (number of
$d/k$-fold covers of such curves), with the same constraint.  A degree
$k$ curve intersects a hyperplane in $k$ points, giving us $k$ choices
for the marking $f(p_i)$; we also have $d/k$ choices for the pre-image
$p_i$ of the covering map, but since we have three markings, these
choices are equivalent (by an automorphism of $\P^1$).  Overall, we
have $ \< H^3 \>_{0,d} = \sum_{k|d} n_k k^3 $.  On the other hand,
this equals $d^3 \< 1 \>_{0,d} = d^3 N_d $ by the divisor axiom, and
we obtain our claim.

\subsection{Gromov-Witten Potentials}

From now on, we restrict ourselves to $X$ being a CY threefold.  The
potential we give here is specifically derived in this context, but
the logics and many of the formulae hold more generally.  It relies on
the convenient feature that $\< 1 \>_{g,d}$ exists for CY threefolds
and that non-zero GW invariants are of the form $ \< \gamma_1 \dots
\gamma_n \>_{0,d} = d_1 \dots d_n ~\< 1 \>_{g,d} $ ({\em i.e.}~all
$\gamma_i$ must be 2-forms).  The letter $d$ now denotes the vector
(of length $h^{1,1}(X)$) representing the class $ \beta = \sum d_i
\gamma^i$ in a basis $\gamma^1, \dots, \gamma^{h^{1,1}}$ of
$H_2(X,\Z)$.

\subsubsection{Summing over Holomorphic Maps}

Given a complexified K\"ahler class $ \gamma=\sum t_i \gamma_i \in
H^2(X,\C) $, with $t_i\in\C$ and $\gamma_i$ forming a basis of
$H^2(X,\Z)$, we would like to build functions $F_g (t_i, \bar{t}_i)$
of the complexified K\"ahler moduli $(t_i, \bar{t}_i)$ using our space
$ \mbar_{g,n}(X,\beta) $.  However, this space carries a dependence on
$ \beta \in H_2(X,\Z) $, and to obtain a dependence on $(t_i,
\bar{t}_i)$ only, we simply sum over all possible such classes $
\beta$.  It turns out that the correct procedure is to rather sum over
all possible maps $f:\Sigma_g \to X$ whose images yield these classes.
Of course the latter sum is much bigger since there are several maps
$f$ (perhaps infinite families) with image in one same class $\beta$.
In fact, summing over all inequivalent maps to $X$ ({\em i.e.}~$
\sum_{f:\Sigma_g \to X} $) amounts to sum over all $ \beta \in H_2$
and weigh each term by the size of the moduli space of maps to
$\beta$.  This weighing factor is our basic Gromov-Witten invariant $
\< 1 \>_{g,\beta}:= N_d^g := $ deg $ [\mbar_{g,0}(X,\beta) ]^{\rm vir}
$.  Thus we can replace $ \sum_{f:\Sigma_g \to X} \dots $ by $
\sum_\beta \< 1 \>_{g,\beta} \dots $ or $ \sum_{d} N_d^g \dots $.

We introduce the notation $ \int_\beta \gamma = \beta \cap \gamma =
\sum d_i t_i =: d \cdot t$ (here $ d \in \N^{h^{1,1}}$ and $ t \in
\C^{h^{1,1}}$).  We furthermore restrict the dependence of $F_g (t_i,
\bar{t}_i)$ on the {\it holomorphic} K\"ahler moduli ($t_i$), {\em
  i.e.}~we demand the maps $f:\Sigma_g \to X$ to be holomorphic.  We
choose to sum over the exponential $e^{d \cdot t}$, and to guarantee a
finite expression we require $\gamma \in$ complexified K\"ahler cone
such that $d \cdot t <0 $.  The function we now introduce is known as
the genus $g$ {\it Gromov-Witten potential} associated to the K\"ahler
class $ \gamma =\sum t_i \gamma_i$:
\begin{equation}  \label{potential}
  \begin{split}
F_g(t_i) &:= \sum_{\substack{\textrm{hol maps} \\ f:\Sigma_g \to X}} 
  \exp{\int_{\Sigma_g} f^*(\gamma)}  
  = \sum_{\substack{\textrm{hol maps} \\ f:\Sigma_g \to X}}
  \exp{\int_{\beta} \gamma} \\ 
  &= \sum_{\beta \in H_2} \< 1\>_{g,\beta} \: e^{\beta \cap \gamma} 
   = \sum_{d} N_d^g ~ e^{d \cdot t} =  \\ 
  &= \< 1\>_{g,0} +  \sum_{\beta \neq 0} 
  \< 1\>_{g,\beta}  ~e^{\beta \cap \gamma} \\
  \end{split}
\end{equation}
where further $ e^{\beta \cap \gamma} = \sum_{n \geq 0} \frac{1}{n!}
~({\beta \cap \gamma})^n = \sum_{n \geq 0} \frac{1}{n!} \sum_{i_1
  \ldots i_n} t_{i_1} \ldots t_{i_n} (\beta \cap \gamma_{i_1}) \dots
(\beta \cap \gamma_{i_n}) $, \\
while $ \< 1\>_{g,\beta} ~(\beta \cap \gamma_{i_1}) \dots (\beta \cap
\gamma_{i_n}) = \< \gamma_{i_1} \ldots \gamma_{i_n} \>_{g,\beta}$ by
the divisor axiom. Thus:
\begin{equation*}
  \begin{split}
F_g(t_i) &= \< 1 \>_{g,0} + \sum_{\beta \neq 0, n \geq 0} \frac{1}{n!}
  \sum_{i_1 \ldots i_n} t_{i_1} \ldots t_{i_n}  \< \gamma_{i_1} \ldots
  \gamma_{i_n} \>_{g,\beta} \\ 
  &= \< 1 \>_{g,0}  +  \sum_{\beta \neq 0, n \geq 0} \frac{1}{n!}
  \< \gamma^n \>_{g,\beta} \\
  &= \< 1 \>_{g,0}  +  \sum_{\beta \neq 0} \< e^{\gamma} \>_{g,\beta}
 \qquad =: F_g^{\rm cl} +F_g^{\rm qu},
  \end{split}
\end{equation*}
where we consider the contribution from constant maps as the {\em
  classical} potential, while all additions for $\beta \neq 0$ are
seen as {\em quantum} corrections.  

We group all potentials into one single {\em full GW potential}
$$
F(t,\lambda) := \sum_{g\geq 0} F_g ~\lambda^{2g-2},
$$
where $\lambda$ is the string coupling constant and plays the role of
formal expansion paramter.

For non-Calabi-Yau target space, the last set of equations are general
enough to define $F_g$, but the $\gamma_i$'s span the whole of $H^*(X,\C)$.  
So much for the general theory.  Let us review some well-known
contributions to the $F_g$.  

\subsubsection{Constant Map Contribution for CY Threefolds}
\label{sec:const-maps} 

The contribution from constant maps ($\beta = 0$) to the full GW
potential is well-known \cite{GeP-98}\cite{KM-94}. It requires the
{\em Hodge bundle} $\E$ for the universal curve $\pi:\mbar_{g,n+1}
\to \mbar_{g,n} $: this is the rank $g$ bundle on $ \mbar_{g,n}$ with
fibre $H^0(\Sigma,\omega_\Sigma)$ over the point
$(\Sigma,p_1,\dots,p_n)$.\footnote{That is, the $g$ independent
  holomorphic differentials on the Riemann surfaces of genus $g$ form
  vector spaces which patch together to form the Hodge bundle.
  Alternatively, $ \E := \pi_* \omega_C$ where $C=\mbar_{g,n+1}
  /\mbar_{g,n}$ is the universal curve and $\omega_C$ the relative
  dualising sheaf.}  Denote its Chern classes by
\underline{$\lambda_i:= c_i(\E)$}.

For constant maps, the splitting $ \mbar_{g,n} (X,0) =
\mbar_{g,n} \times X $ is accompanied by the splitting of the
obstruction sheaf Ob $= \E^* \boxtimes T_X$.  So from
(\ref{vir-fund-class-ns}): 
$$
 [ \mbar_{g,n}(X,0) ]^\virt = [ \mbar_{g,n} \times X ] \cap e(\E^*
 \boxtimes T_X).
$$ 
With this split, the Euler character of the obstruction bundle is
 easy to calculate \cite{GeP-98}:
$$ 
e(\E^* \boxtimes T_X) = \frac{(-1)^g}{2} ~( c_3(X) -c_2(X)~c_1(X) )
~\lambda_{g-1}^3   , \hspace{2cm} g \geq 2,
$$ and in the case of a Calabi-Yau threefold the expression reduces to
$ \frac{(-1)^g}{2} ~c_3(X) ~\lambda_{g-1}^3 $.  Due to the occurrence
of $ c_3(X) $, an Euler class, we cannot allow for $ \gamma_i$'s to be
integrated over $X$, so the only non-zero GW invariants for $\beta
=0$, $g \geq 2$ has $n=0$:\footnote{ 
  The integral in $ \int_{\mbar_{g,0}} \lambda_{g-1}^3 $ is our first
  (and only) example of {\it Hodge integrals}, {\em i.e.}~integrals over $
  [\mbar_{g,n}(X,\beta)]^\virt $ that include $\psi$ and
  $\lambda$-classes: $$ \int_{ [\mbar_{g,n}(X,\beta)]^\virt }
  \prod_{i=1}^n \psi_i^{a_i} \cup {\rm ev}_i^* (\gamma_i) \cup
  \prod_{j=1}^g \lambda_j^{b_j} $$ where the $\gamma_i \in H^*(X,\Q)$,
  {\em i.e.}~not necessarily 2-forms, and the $\psi_i $ are cotangent line
  classes over $ \mbar_{g,n}(X,\beta) $.  Without $\lambda$-classes we
  have {\it gravitational descendants}, and with no $\psi$-classes
  either we have {\it primary} GW invariants.  A theorem of
  \cite{FP-98} states that Hodge integrals over $
  [\mbar_{g,n}(X,\beta)]^\virt $ can be uniquely reconstructed from
  the set of descendent integrals.  In genus 0 or 1, the latter can
  even be expressed in terms of primary GW
invariants \cite{KM-94}. \\ 
Note that in our case, the $ \lambda$-class was hidden in the virtual
fundamental class and not included as an extra parameter (as in
genuine Hodge integrals over $ \mbar_{g,n}(X,\beta) $).  Since the
Hodge class $ \lambda_{g-1}^3 $ already matches the dimension of $
\mbar_{g,0} $, we cannot allow for extra classes; and anyway $n=0$
means we have no $\psi$ classes in the above constant term.}
\begin{equation}  \label{constant-map}
\< 1 \>_{g,0} =  \frac{(-1)^g}{2} \chi(X) \int_{\mbar_{g,0}}
\lambda_{g-1}^3  = (-1)^g \frac{\chi(X)}{2} \frac{B_{2g}}{2g
  (2g-2)!} \zeta(3-2g),
\end{equation}
where $B_{2g}$ are Bernoulli numbers\footnote{Bernoulli numbers are
  defined by $ \sum_{0} B_k \frac{x^k}{k!}:= \frac{x}{e^x - 1}$; they
  satisfy $B_{2k+1}=0$ except $B_1=-\half$, and $B_{2k} = (-1)^{k+1}
  {2(2k)! \over (2\pi)^{2k}} ~\zeta(2k) = -2k ~\zeta(1-2k)$.}.  The
first step was reached in \cite{BCOV-94} and the last step was
computed in \cite{FP-98}.  Hence, if we formally set $ \< 1 \>_{0,0}
:= - \frac{\chi}{2} \zeta(3)$ and $ \< 1 \>_{1,0} := - \frac{\chi}{2}
\zeta(1) / 12 $ for the genus 0 and 1 contributions, and using the
expansion (\ref{inverse-sine}), we obtain the following constant map
contributions to the full GW potential:
\begin{equation} \label{F_const}
    F_{\textrm{const}} := \sum_{g \geq 0} \lambda^{2g-2} \< 1 \>_{g,0}
= - \frac{\chi(X)}{2} \sum_{k=1} \frac{1}{k} \frac{1}{(2 \sin \frac{k
    \lambda}{2})^2}  
= - \frac{\chi(X)}{2} \sum_{n=1} \log(1-e^{\pm i \lambda n})^n.
\end{equation}
This is the generating function for solid partitions ($3d$
partitions), as proven originally in \cite{M-1915} and related in this
context to melting crystals by \cite{ORV-03}.

For the \underline{$g=0$} case properly, we have $\E=0$, so the
obstruction bundle is trivial and the virtual fundamental class does
not furnish us with a $ \lambda$-class to integrate over $
\mbar_{0,n}$ (which is $n-3$ dimensional).  Hence $ e(\E^* \boxtimes
T_X) = 1$ and $ \< 1\>_{0,0} = 0$, and the only non-zero primary GW
invariant occurs for $n=3$, as mentioned in (\ref{const-map-axiom}):
$\< \gamma_1 \gamma_2 \gamma_3 \>_{0,0} = \int_X \gamma_1 \cup
\gamma_2 \cup \gamma_3$.

For \underline{$g=1$}, we have \cite{GeP-98} $ e(\E^* \boxtimes T_X) =
c_3(X) - c_2(X) \lambda_1 $, so the only non-zero primary GW invariant
occurs for $n=1$:
\begin{equation}  \label{const-map-ell}
 \< \gamma_{i} \>_{1,0} =  [\mbar_{1,1}(X,0)]^\virt 
= - \int_{X} c_2(X) \cup \gamma_{i} \int_{\mbar_{1,1}} \lambda_1
= - \frac{1}{24} \Big( c_2(X), \gamma_{i} \Big) 
\end{equation}
Note that $ \< 1\>_{1,0} $ would be acceptable from the dimensional point
of view (expecting $ \int_{X} c_3(X) \int_{\mbar_{1,0}} 1 $, similar to
the $g=0, n=3$ term), but we rule this case out since it is unstable
($ 2g-2+n \leq 0$). 

\subsection{Multicovers and Known Cases of Excess Intersection}

Having defined the GW invariants $N_d^g$, we now turn to the question
of how a given image curve $C$ in the CY threefold $X$ contributes to
them.  This question will ultimately lead us to the physical approach
of string theory.  

\subsubsection{Multicover Contributions} \label{sec:multicover-contributions}

So far we have only mentioned the {\it degree} $\beta$ (or $d$) of our
maps, carefully avoiding the nature of the image curve.  However, the
latter also has a genus which we denote by $r$, and the
Riemann-Hurwitz theorem tells us that a cover of such a curve can only
come from a Riemann surface $\Sigma_g$ with $g \geq r$.  So $N_d^g$
receives contributions from image curves $C_r$ with genus $r$ not
greater than $g$.  Moreover, the degree of the curve need not equal
$\beta$, as $N_d^g$ also has contributions from $k$-fold
covers of curves of degree $\beta /k$ if $\beta /k \in H_2(X,\Z)$.
Overall, we wonder what is the contribution to $N_d^g$ from the map
$ f:\Sigma_g \rightarrow C_r $ of degree $k$.  The additional
distinction of the {\it genus} of the image curve is typical of the
topological string theory approach, where this number has the physical
meaning of the $ SU(2)_L $ content of the spectrum of M2-branes
wrapped on the curve (see below).

Let $C$ be an image curve of degree $\beta \neq 0$ and arithmetic
genus $r$.  Define $h$ by $g=:r+h$.  Then $C$ contributes to the genus
$g$ degree $k \beta$ basic GW invariant via $ \mbar_{g,0}(X,k \beta)$,
{\em i.e.}~via a $k$-fold cover of $C$ (sometimes called degenerate
contributions, though we shall keep this term for stricter cases).
Let us denote $C$'s contribution by $C_r(h,k)$.\footnote{A good
  definition of this excess contribution, avoiding {\it rigidity}
  problems is the following: 
  $$
  C_r(h,k) := \int_{[\mbar_{r+h,0} (C,k \beta)]^\virt} e(R^{\bullet
  } \pi_* ~{\rm ev}_1^* ~(\CO_C \oplus K_C)[1]).
  $$
  where $K_C $ is the canonical sheaf.}
Then $N_d^g$ gets a contribution only from a curve $C$ of degree $d$,
but also from multicovers of curves of smaller degree and lower
genus:
\begin{equation}  \label{Ngd}
  N_d^g =: \sum_{\substack{k|d \\ r \leq g}} C_r(g-r,k) \ n^r_{d/k} 
\end{equation}
where $k|d$ means $k|d_i \; \forall i$.  This excess intersection
formula is a generalisation of the previous one for rational curves
(\ref{quintic-rational}).  In this way of writing, $n^r_{d/k}$ takes
the meaning of ``number of holomorphic curves'' of arithmetic genus
$r$ and degree $d/k$, since each such curve contributes an amount of
$C_r(g-r,k)$.  Equation (\ref{Ngd}) actually defines the $n^r_d$, and
we suggest to call them the ``virtual number of curves''.  One expects
them to be rational numbers for two reasons: first because $N^g_d$ and
$C_r(h,k)$ are so; secondly because ``counting'' curves that are not
isolated (occur in families as is often the case) involves integrals
of Euler classes over such families and the latter Euler numbers are
typically rational.  Thus one is startled to discover that in all
known examples, the $n_d^r$ computed from known $N_d^g$ and
$C_r(g-r,k)$ turned out to be integers.  This has remained a mystery
ever since GW theory was developed.

We now introduce the so-called {\em BPS-invariants} $\tilde{n}^r_d$,
which differ only slightly\footnote{The only difference occurs at
  genus 1 ($ \tilde{n}_d^1 = \sum_{k|d} n^1_k $, see section
  \ref{sec:subtleties}) and genus 2 ($ n^2_d = \sum_{k|d} {d\over k}
  \tilde n^2_k $
).  These subtleties are rather
  scholastic and we encourage the reader to overlook them at a first
  reading.}  from the above $n^r_d$.  They have their origin in the
physical approach (see below) via the D-brane moduli space $\CM$.  If
the latter is smooth, we can define them by
\begin{equation}  \label{physical-n^r_d}
\tilde{n}^r_d := \int_{\CM} e(T^*\CM) = (-1)^{\textrm{dim}\CM} e(\CM).
\end{equation}
We will see shortly a more usual way to define them (see eqn
(\ref{F})).  It is useful to relate them to the $N_d^g$ via a similar
version of (\ref{Ngd}):
$$
  N_d^g =: \sum_{\substack{k|d \\ r \leq g}} \tilde{C}_r(g-r,k) \
  \tilde{n}^r_{d/k}
$$
which may be seen this time as defining the $\tilde{C}_r(h,k)$.  We
will see that the $\tilde{n}^r_d$ are integers iff the $n^r_d$ are so.
The {\em Gopakumar-Vafa conjecture} states that the $\tilde{n}^r_d$
are indeed integers.

\subsubsection{Degree-Genus Relationship} \label{degree-genus}

One will expect the $n_d^r$ to be non-zero only for specific
combinations of $r$ and $d$, since the arithmetic genus depends on the
degree of the curve.  For example, if the Calabi-Yau is a
local\footnote{We call {\em local CY threefold} the total space $X$ of
  a fibration $ \CO (K_B) \to B$ with fibre the canonical bundle of a
  surface $B$. This is non-compact but locally CY since $c_1(X)
  =c_1(\CO(K_B) \o TB) = c_1(\CO(K_B)) + c_1(B) =0$.  The line bundles
  $\CO_{\P^d}(n) :=\CO_{\P^d}(nH)$, for $n\in\Z$ and $H$ the divisor
  class of $\P^d$, are those whose transition functions are locally
  $(x_\alpha/x_\beta)^n$ and whose sections are the degree $n$
  homogeneous polynomials (also called {\em twisted sheaves of Serre}
  in \cite{H-77}).  Similarly, the base could rather be 1-dimensional
  and the fibre a rank two bundle, as in $ \CO (-1) \oplus \CO (-1)
  \to \P^1$.}  $\P^2$, {\em i.e.}~the fibration $\CO(-3) \to \P^2$ with
fibre the line bundle $\CO(-3)$, then a curve of degree $d$ can only
sit in the plane and will have the usual arithmetic genus of $
r=(d-1)(d-2)/2$.  One can add $\delta $ nodes to the curve and thus
lower the genus to obtain a geometric genus of $ r-\delta$; this will
contribute to the invariant $ n^{r-\delta}_d$.  Thus, from now on, we
admit singular image curves and the superscript in $n^r_d$ will
actually denote the {\em geometric} genus.
Another example is the local $\P^1 \times
\P^1$ case, where a curve of bidegree $ (a,b)$ with respect to the two
$ \P^1$'s has a total degree $d=a+b$ and arithmetic genus
$(a-1)(b-1)$.

A third example \cite{KKV-99} of a local CY case, is given by choosing
the base to be a del Pezzo surface $E_n$ ({\em i.e.}~the blow-up of
$\P^2 $ in $n$ points): a curve in the base is described by
multidegrees $d=(a; b_1 \dots b_n)$ corresponding to the class $a H -
\sum b_i e_i $ ($H$ is the hyperplane class and $e_i$ are the
exceptional divisors of the blow-ups in $n$ points).  Its genus is
given by (see {\em e.g.}  section~4.4.1 of \cite{G-04})
$$ 
r=\frac{(a-1)(a-2)}{ 2}-\sum_{i=1}^n \frac{b_i(b_i-1)}{2}
$$
and its overall degree is given by its intersection number with the
anticanonical class $-K_{E_n}=3H-\sum_{i=1}^n e_i$: 
$$
d=-K_{E_n}\cdot (a; b_1 \dots b_n)= 3 a-\sum_{i=1}^n b_i ~,
$$ 
where we have exploited $H^2= 1$ and $ e_i e_j = - \delta_{ij} $.  We
use this new integer $d$ to label the $n_d^r$.  

So these examples reflect the general pattern that one finds a curve
of given genus only at sufficiently large degree.  This is not to say
that this curve will be no candidate for $n_d^r$ with lower $r$:
adding $\delta$ nodes to the curve lowers its geometric genus to
$r-\delta$.  In that way, singular curves are taken into account for
the invariants $n_d^{r-\delta}$ for $\delta=1,\ldots,r$.  So the
numbers $ n^r_d$ only vanish if $r$ is too large compared to $d$.
Determining the maximal genus of a curve of given degree $d$ in a
particular space is the subject of Castelnuovo theory.

\subsubsection{Known $C_r(h,k)$ and a Physical Assumption}

The $C_r(h,k)$ have been computed for $r=0,1$:

For \underline{rational curves} ($r=0$), \cite{FP-98} showed that for
CY threefolds: 
$$
\sum_{h \geq 0} C_0(h,1) ~\lambda^{2h} =
\left( \frac{\sin{\lambda/2}}{\lambda/2} \right)^{-2} 
$$
\begin{equation}  \label{rational}
  C_0(h,k)= C_0(h,1) ~\frac{1}{k^{3-2h}} = C_0(h,1) ~\frac{1}{k^{3-2g}}
\end{equation}
yielding in closed form: 
$$  C_0(h,k) =  \left\{
\begin{array}{ll}
k^{2g-3} ~|B_{2h}|/ 2h (2h-2)!  & h \geq 2 \\
1/k^3 & h=0 \\
1/(12 k) & h=1
\end{array} \right. 
$$ 
The $B_{k}$ are the most fashionable version of Bernoulli numbers,
defined under (\ref{constant-map}).  The cases $h=0,1$ were long
known.

For \underline{elliptic curves} ($r=1$), the Riemann-Hurwitz formula
allows for unbranched multicovers from other elliptic curves ({\em i.e.}
$g=1$, {\em i.e.}~$h=0$).  \cite{P-98}(v.2 !) showed:
\begin{equation}  \label{elliptic}
C_1(0,k) = \frac{\sigma_1(k)}{k} = \sum_{n|k} \frac{1}{n} \quad
\textrm{and} \quad C_1(h,k) = 0 \qquad (h > 0)  
\end{equation}
The last equation claims that elliptic curves only contribute to
$F_1$, although in principle the Riemann-Hurwitz formula admits
branched covers from higher genus curves.\footnote{In \cite{BCOV-94} 
  this was argued to be due to the fact that the moduli space
  $\mbar_{1+h,0}(X,\beta)$ does not contain a factor of a torus.  This
  absence of flat torus is crucial since factors of non-zero curvature in the
  action are needed to absorb the fermion zero modes.}  

Generalising to \underline{higher genera}, \cite{P-98} proved:
\begin{equation}  \label{crh1}
\sum_{h \geq 0} C_r(h,1) ~\lambda^{2h} =  \left( \frac{\sin
  \lambda/2}{\lambda/2} \right)^{2r-2}.
\end{equation}
We assume that the $\tilde{C}_r(h,k)$ enjoy the same properties,
except for the first part of (\ref{elliptic}):
\begin{equation}  \label{elliptic2}
\tilde{C}_1(0,k) = \frac{1}{k}  
\end{equation}
With this small change done, we now would like to \underline{assume} that 
(\ref{rational}) also holds at higher genera, {\em i.e.}~
\begin{equation}  \label{assumption}
\tilde{C}_r(h,k) = \tilde{C}_r(h,1) ~\frac{1}{k^{3-2(r+h)}}
\end{equation}
which entails that $\tilde{C}_g(0,k) =k^{2g-3}$.  Note that the
$\tilde{C}_r(h,k)$ and the $C_r(h,k)$ only differ for $k>1$.  The
assumption (\ref{assumption}) cannot hold for the $C_r(h,k)$
themselves as it would suggest that multicovers of a genus $g$ curve
by another genus $g$ curve yield a non-zero contribution, in
contradiction to the
Riemann-Hurwitz formula which prohibits such multicovers.  
In the D-brane or BPS picture, we face no such embarrassment, as 
D-branes are more than just curves; in fact they carry gauge
bundles and are best treated as sheaves on the algebraic geometric
side. In any case, (\ref{assumption}) is rather an assumption
cherished by physicists.  We will see why this is a powerful
assumption; it will enable us to derive now the expression by which
the $\tilde{n}^r_d$ are usually defined -- eqn (\ref{F}).

\subsection{Explicit Forms of $F$ and $F_g$ for CY Threefolds}
\label{sec:explicit-forms-F}

This section is devoted to algebraic manipulations of sums and
products in order to arrive at elegant versions for the GW potentials
on CY threefolds.

\subsubsection{Evaluating $F$ in Terms of $\tilde{n}^r_d$}

We will combine all GW potentials $F_g$ into a single generating
function $ F:= \sum_0 \lambda^{2g-2} F_g $; then the latter result
(\ref{crh1}) and assumption (\ref{assumption}) enter in the last step
of the next calculation.  For simplicity, we drop the constant map
terms ($\beta =0$) from the potentials and call the new potentials
\underline{$ \tilde{F}_g$ and $ \tilde{F}$}.  We also introduce $q^d
:= e^{t \cdot d} $.
\begin{equation}  \label{F_g}
\tilde{F}_g (t_i) = \sum_{d>0} N_d^g \ q^d 
= \sum_{d, \ k|d, ~r \leq g} \tilde{C}_r(g-r,k) \ \tilde{n}^r_{d/k} \ q^d
= \sum_{d, k>0, ~r \leq g} \tilde{C}_r(g-r,k) \ \tilde{n}^r_d \ q^{dk}
\end{equation}
\begin{equation*}
  \begin{split}
    \tilde{F}(t_i) &= \sum_{g=0} \lambda^{2g-2} \tilde{F}_g 
    = \sum_{\substack{r,h\geq 0 \\ d,k>0}} \lambda^{2(r+h)-2} \
    \tilde{C}_r(h,k) \ \tilde{n}^r_d \ q^{d k} \\
    &= \sum_{r \geq 0, ~d>0} \tilde{n}^r_d \ \lambda^{2r-2} \sum_{k>0} q^{d k}
    \sum_{h\geq 0} \tilde{C}_r(h,k) \ \lambda^{2h} \\
    &= \sum_{r \geq 0, ~d>0} \tilde{n}^r_d \ \lambda^{2r-2} \sum_{k>0} q^{d
    k} k^{2r-3} \sum_{h\geq 0} \tilde{C}_r(h,1) \ (k \lambda)^{2h} \\
    &= \sum_{r \geq 0, ~d>0} \tilde{n}^r_d \ \lambda^{2r-2} \sum_{k>0}  
    \frac{1}{k} \left(\frac{\sin \frac{k \lambda}{2}}{\lambda/2}
    \right)^{2r-2} \ q^{d k}
  \end{split}
\end{equation*}
Hence
\begin{equation}  \label{F}
  \tilde{F}(t_i) =\sum_{\substack{r\geq 0\\ d,k>0}} \tilde{n}^r_d ~\frac{1}{k}
  \left(2 \sin \frac{k \lambda}{2} \right)^{2r-2} q^{d k}
\end{equation}
This last expression was obtained by the physical approach of
\cite{GV2-98} where the $\tilde{n}^r_d$ had a more pertinent meaning
(see section~\ref{sec:physical-meaning}): the number of BPS states of
charge $d$ and $SU(2)_L$ content $r$ in the M-theory compactification
on the Calabi-Yau $X$.  From a mathematical point of view, one may
consider (\ref{F}) as giving an alternative definition of the
$\tilde{n}^r_d$, avoiding physical quantities.

\subsubsection{Evaluating $F_g$ in Terms of Polylogs ${\rm Li}_{3-2g}(q^d)$}

We shall now rewrite $F_g$ in terms of polylogarithms and present some
explicit examples of GW potentials or GW invariants like the local
$\P^1$ case, an infinite product expression, or the 3-point
function.  For convenience we recall the following expansions:
\begin{equation}  \label{inverse-sine}
\left(2 \sin \frac{k \lambda}{2} \right)^{-2} =
   \frac{1}{(k\lambda)^2} + \frac{1}{12} + \frac{(k\lambda)^2}{240} +
   \ldots + \frac{|B_{2n}|}{2n(2n-2)!} (k\lambda)^{2n-2} + \ldots 
\end{equation}
$$
 \left(2 \sin \frac{k \lambda}{2} \right)^2 = (k\lambda)^2 -
   \frac{(k\lambda)^4}{12} + \ldots + (-1)^{n+1} 2
   \frac{(k\lambda)^{2n}}{(2n)!} + \ldots
$$
This allows us to express the $\tilde{F}_g$ in terms of the
$\tilde{n}^r_d$.  For low genus we have:
\begin{equation}\label{low-genus-F_g}
  \tilde{F}_0 = \sum_{d>0}\tilde{n}_d^0 \ \Li_3(q^d) \qquad
\tilde{F}_1 = \sum_{d>0} \left( \frac{1}{12} \tilde{n}_d^0 +
  \tilde{n}_d^1 \right) \Li_1(q^d) 
\end{equation}
\begin{equation}  \label{F_2}
  \tilde{F}_2 = \sum_{d>0} \left( \frac{1}{240} \tilde{n}_d^0 +
    \tilde{n}_d^2 \right) \Li_{-1}(q^d)  
\end{equation}
and in general ($g\geq 2$):
\begin{equation}  \label{F_g-expl}
  \tilde{F}_g (t_i) = \sum_{d>0} \left( \frac{|B_{2g}|}{2g(2g-2)!} \ 
  \tilde{n}^0_d + \frac{2 (-1)^g}{(2g-2)!} \ \tilde{n}^2_d + \ldots +
  \tilde{n}^g_d \right) \Li_{3-2g}(q^d)
\end{equation}
The polylogarithms are defined by $ \Li_g (q) := \sum_{1}
\frac{q^n}{n^g} $ and satisfy 
$$ 
\Li_1 (q)= -\log(1-q), \quad \Li_0 (q) = \frac{q}{1-q},\quad
\Li_{-1} (q)=  \frac{q}{(1-q)^2}
$$
and in general: $ q \p_q ~\Li_{g} = \Li_{g-1} $. 

Of course these relations for $\tilde{F}_g$ could have been directly
obtained via (\ref{crh1}), (\ref{assumption}) and (\ref{F_g}).  In any
case, the polylogarithms occur due to the multicovers, {\em i.e.}~the sum over
$k$.  It should be emphasised that we could not replace
$\tilde{n}^r_d$ by $n^r_d$ in $\tilde{F} $ and $\tilde{F_g}$, as the
assumption (\ref{assumption}) would not hold for the algebro-geometric
$C_r(h,k)$.  The result (\ref{F}) was first derived by
\cite{GV2-98}\cite{KKV-99} from a physical approach (see
section~\ref{sec:schwinger-one-loop}). 

\subsubsection{Example: the Local $\P^1$}

An easy example of generating function for GW potentials is the local
${\P^1}$ case ({\em i.e.}~non-compact CY space), the rank two concave
bundle $ \mathcal{O}(-1) \oplus \mathcal{O}(-1) \longrightarrow \P^1 $.  
Here the conifold geometry forbids positive genus curves.
The base $\P^1$ is the only rational curve, so all $n^r_d$ vanish
except $n^0_1 = 1$ (since $ h^2(\P^1) = 1 $, $d$ is merely an
integer).  From (\ref{F_g-expl}) we then have:
\begin{equation}\label{local-P1}
  \begin{split}
    \tilde{F}_{\P^1} &= \sum_{\geq 0} \lambda^{2g-2} \tilde{F_g} =
    \lambda^{-2} \Li_3(q) + \sum_{g \geq 1} \lambda^{2g-2} \ 
    \frac{|B_{2g}|}{2g(2g-2)!} \ \Li_{3-2g}(q) \\
    &= \sum_{l \geq 1} \frac{q^l}{l} \left( (l \lambda)^{-2} +
    \sum_{g \geq 1} \frac{|B_{2g}|}{2g(2g-2)!} ~(l \lambda)^{2g-2} \right) \\
    &= \sum_{l \geq 1} \frac{q^l}{l}\frac{1}{(2 \sin \frac{l \lambda}{2} )^2} 
    \qquad =  \sum_{n \geq 1} \log (1-e^{\pm i\lambda n} q)^n 
  \end{split}
\end{equation}
And adding the constant map contribution (\ref{F_const}) with $ \chi
(\P^1)=2$:  
$$
F_{\P^1} = \sum_{n \geq 1} \log [ (1-e^{\lambda n})^{-n}
(1-e^{\lambda n} q)^n ],
$$
where we take the freedom to treat the string coupling $\lambda$ modulo
$\pm i$. 

\subsubsection{Toda Equation for the Local $\P^1$}

This is in fact not the full story.  Physical considerations require
to add some `classical terms' to this `quantum' piece.  The classical
part only occurs at genus 0 and 1 ({\em i.e.}~powers $\lambda^{-2}$ and
$\lambda^0$) and is a polynomial in the modulus $t$: 
$$
F_{\textrm{cl}}(t) = \lambda^{-2} \Big( - \zeta (3) + \frac{\pi^2}{6} t +i
(m+\frac{1}{4}) \pi t^2 - \frac{1}{12} t^3 \Big) + \frac{1}{24} t
$$  
Under shifts of $ t \rightarrow t \pm \lambda$, this classical
part behaves as:
$$
F_{\textrm{cl}} (t \pm \lambda) = F_{\textrm{cl}} (t) +
\lambda^{-2} \Big( \mp \frac{1}{4} t^2 \lambda - \frac{1}{4} t
\lambda^2 \pm \frac{1}{12} \lambda^3 +i(m+\frac{1}{4}) \pi
(\pm 2 t \lambda + \lambda^2) \pm \frac{\pi^2}{6} \lambda
\Big) \pm \frac{\lambda}{24},
$$
so that 
$$ F_{\textrm{cl}} (t+\lambda) + F_{\textrm{cl}} (t-\lambda) - 2
F_{\textrm{cl}}(t) = - \frac{t}{2} + i (2m +\frac{1}{2})\pi. 
$$
Or in terms of the partition function $Z=e^F$:
$$
\frac{Z_{\textrm{cl}} (t+\lambda) Z_{\textrm{cl}} (t-\lambda)}
{Z_{\textrm{cl}}(t)^2} = e^{-\frac{t}{2}+i\frac{\pi}{2}}.
$$
Why are we particularly interested in this last combination~? Well,
the left hand side is just the homogeneous part of the Toda equation;
and we now would like to see how the `quantum' piece of our local
$\P^1$ GW potential behaves.  From the above product form of
$F_{\P^1}$ and using $e^{\lambda n} e^{t\pm \lambda} = e^{\lambda
  (n+1)} q $, we readily obtain:
$$
Z_{\textrm{qu}}(t+\lambda) = Z_{\textrm{qu}}(t) \prod_{n=1}(1-e^{n
  \lambda} q)^{-1}, \qquad Z_{\textrm{qu}}(t-\lambda) =
Z_{\textrm{qu}}(t) \prod_{n=0}(1-e^{n \lambda}q),
$$
so that
$$
\frac{Z_{\textrm{qu}}(t+\lambda)Z_{\textrm{qu}}(t-\lambda)}
{Z_{\textrm{qu}}(t)^2} = (1-q),
$$
and the similar expression for the full (classical + quantum) change
in the GW partition function of the local $\P^1$ reads:
$$
\Delta Z = q^{-1/2} i (1-q) = -2i \sinh\frac{t}{2}.
$$
Thus the partition function of the local $\P^1$ satisfies the
homogeneous Toda equation up to this inhomogeneity.

\subsubsection{The Product Expression}

If we had not restricted to the local $\P^1$ case and were
interested in the part of the GW potential coming only from constant maps
and rational curves ($r=0$) on a general CY threefold, we would simply put
back the $\tilde{n}^0_d$ in the above (\ref{local-P1}) and obtain:
\begin{equation*} 
  \begin{split} 
    F_{_{D0-D2}} &= \sum_{n \geq 1} \log \Big[ (1-e^{\lambda
      n})^{-n \frac{\chi}{2} } \prod_{d>0} (1-e^{\lambda n} q^d)^{n
      \tilde{n}_d^0} \Big] \\ 
    &= \log \prod_{\substack{n \geq 1 \\ d \geq 0}} (1-e^{\lambda n}
    q^d)^{n \tilde{n}_d^0} 
  \end{split}
\end{equation*}  
setting formally $ \tilde{n}_0^0 := -\frac{\chi}{2}$.  This is just the
Gopakumar-Vafa result \cite{GV1-98} for contributions from constant plus
rational maps ({\em i.e.}~D0- and D2-branes).

In the same vein, we can rewrite the full GW potential (\ref{F}) in
product form by Fourier expanding the sine term:
\begin{equation*}
  \begin{split}
    (2 \sin x/2 )^{2r-2} &= \sum_{|j| \leq
     r-1} (-1)^j {2r-2 \choose r-1+j} e^{ijx} \qquad
   =: \sum_{|j| \leq r-1} c_{r,j} \ e^{ijx} \qquad
   \textrm{for} \qquad r \neq 0,1,
  \end{split}
\end{equation*}
while the sum is infinite in the $r=0$ case and $c_{0,j} = -j $:
~~$ (2 \sin x/2 )^{-2} = -\sum_{j \geq 1} j \ e^{\pm ijx} $. Then
\begin{equation*}
  \begin{split}
  \tilde{F} &= \sum_{\substack{r \geq 0 \\ d,k>0}} \tilde{n}^r_d \ \frac{1}{k}
  \left(2 \sin \frac{k \lambda}{2} \right)^{2r-2} q^{d k}  \quad 
  = \sum_{\substack{d>0 \\ j \in \Z}} \big( -\sum_{r \geq 0}
  c_{r,j} \ \tilde{n}^r_d \big) \log (1-e^{i \lambda j} q^d) \\
  &= \log \prod_{\substack{d>0 \\ j \in \Z}} (1-e^{i \lambda j}
  q^d)^{M_{d,j}},
  \end{split}
\end{equation*}
where we have set $ M_{d,j} := - \sum_{r \geq 0} c_{r,j} \ \tilde{n}^r_d
$.  In subsection \ref{degree-genus}, we argued that the $
\tilde{n}^r_d $ vanish if $r$ is too large compared to $d$; that is,
for fixed $d$, there are only a finite number of $r$ to take into
account, say $r < r_0(d)$.  This is why the sum for the $M_{d,j}$ is a
finite one.  Moreover, except at $r=0$, the sum over $k$ is finite, so
for fixed d one could write $ k < k_0(d) $; but the $r=0$ case makes
the sum over $j$ actually infinite.

As in the local $\P^1$ case, we can incorporate the constant map
contribution (\ref{F_const}) with $ M_{0,j} = -j {\chi \over 2}$:
$$
F = \log \prod_{\substack{d \geq 0 \\ j \in \Z}} (1-e^{i \lambda j}
  q^d)^{M_{d,j}}.
$$

The motivation behind this alternative expression for the GW potential
will appear later when we shall try to relate this product to a
modular product \`a la Borcherds.  Such infinite products exhibit
interesting symmetries, including automorphic properties.

Unlike the local $\P^1$ case, this very general product does not
satisfy the homogeneous Toda equation up to a small inhomogeneity.
This is due to the complicated nature of the powers $M_{d,j}$.  

\subsubsection{The Three-Point Function}

To conclude this subsection, we give the Yukawa coupling or
three-point function, which is the A-model correlation function for
the quintic threefold in the context of Mirror Symmetry.  Since
$h^{1,1} = 1$ for this CY manifold, there is only one K\"ahler modulus
$t_1$, one K\"ahler class $\gamma_1 =H$ (the hyperplane class), and
the degree $d$ of the image curve is simply an integer :
\begin{equation*}
  \begin{split}    
    \< \gamma_1 \gamma_1 \gamma_1 \>_0 &:= \sum_{\beta} \<
    \gamma_1^3 \>_{0,\beta} \ q^\beta 
    = \sum_{d>0} N_d \ d^3 \ q^d \\
    &= \p_{t_1}^3 \tilde{F_0} =: \Phi_{111} \\
    &= \sum_{d>0} \tilde{n}^0_d \ d^3  \ \Li_0(q^d) = \sum_d
    \big( \sum_{k|d} k^3 \tilde{n}^0_k \big) \ q^d 
  \end{split}
\end{equation*}
where in the third step we have used the divisor axiom for
GW invariants: $ \< \gamma_1^3 \>_{0,\beta} = \< 1 \>_{0,\beta}
\big( \int_{\beta} \gamma_1 \big)^3 =N_d ~d^3$.  From this -- or
directly from the expression for $\tilde{F_0}$ -- we obtain the genus 0
GW invariant (\ref{quintic-rational}): $N_d = \sum_{k|d} k^{-3}
~\tilde{n}^0_{d/k}$. 

\subsubsection{Subtleties Around $n^r_d$ and $\tilde{n}^r_d$}
\label{sec:subtleties}  


In \cite{BCOV-94}, the contribution from genus $2$ curves to
$\tilde{F}_2$ was found to be $ \sum_{d} n^2_d ~q^d $, while
(\ref{F_2}) purports $ \sum_{d} n^2_d$ Li$_{-1}(q^d) = \sum_{d,k}
\tilde{n}^2_d ~k ~q^{dk} $, suggesting that $ n^2_d = \sum_{k|d}
{d\over k} \tilde n^2_k $.  This demonstrates that the $\tilde{n}^r_d$
in general do not have the meaning of `counting curves'.  In fact,
they rather count BPS states, and the extra sum over $k$ stands for
multicovers of embedded D-branes.   For instance, \cite{MM-98} 
have taken \cite{BCOV-94}'s result for $\tilde{F}_2 $ when extracting
the number of genus $2$ curves from a similar expression derived by
entirely different methods (see heterotic treatment below).

The expression (\ref{F}) reflects also the fact that BPS states at
genus 1 do not have any bubblings to higher genera, {\em
  i.e.}~$\tilde{n}^1_d$ only appears in $\tilde{F}_1$ (since $r=1$
implies zeroth power of $\lambda$), see also (\ref{F_g-expl}).  From
the polylog expression (\ref{low-genus-F_g}) for $\tilde{F}_1$, we
recover
$$
N_d^1= \frac{1}{d} \sum_{k|d} \big( \frac{1}{12} ~k ~\tilde{n}^0_k +
k ~\tilde n^1_k \big) = \sum_{k|d} \big( \frac{1}{12~k} ~\tilde{n}^0_{d/k} +
{1\over k} ~\tilde n^1_{d/k} \big)
$$
and (of course) the second part of (\ref{elliptic}) and (\ref{elliptic2}).

If we opt to express $\tilde{F}_1$ in terms of $ n_d^1 $ rather than $
\tilde n_d^1 $ , the choice \underline{$ \tilde{n}_d^1 = \sum_{k|d}
  n^1_k $} will prove propitious:
$$ 
\tilde{F}_1 = \sum_{d>0} N_d^1 \ q^d = \sum_{d>0} \left( \frac{1}{12}
\tilde{n}_d^0 \ \Li_1(q^d) - n_d^1 \ \log P(q^d) \right) ,
$$
which is the form appearing in \cite{BCOV-93} 
(with $ n^1_d$ counting {\em curves}), wherein $ P(x) := \prod 
(1-x^{n}) $ and
$$
- \log P(q^d) = \sum_{n=1} \Li_1 (q^{dn}) 
= \Li_1 \left( \frac{1}{1-q^d} \right) 
=\sum_{n,k \geq 1} \frac{q^{dnk}}{k}.
$$ 
Whence the expression for the genus 1 basic GW invariant in terms of
(geometrical) instanton invariants: 
$$ 
N_d^1= \frac{1}{d} \sum_{k|d} \big(
\frac{1}{12} \ k \ n^0_k + \sigma (\textstyle{d\over k}) \ k \ n^1_k
\big)  
$$
with $\sigma(n):=\sum_{k|n} k $ and with no difference between
$\tilde{n}^0_d$ and $n^0_d$.

So the difference between the physical invariants $ \tilde{n}_d^1$ and
the geometrical invariants $n_k^1 $ is that when using the latter, we
have an extra $ \sum_{n}$ in $\tilde{F}_1$.  From the M-theory
perspective, this is due to bound states of $n$ M2-branes wrapped on
the torus $ \Sigma_1$, being indecomposable stable $U(n)$ bundles over
the torus.  This is in addition to the possibility for a single
M2-brane to wrap the torus in a $k$-fold way.  
These bound states solve a puzzle that occurred on the last page of
\cite{GV2-98}, where the M-theory approach seemed to disagree with
previous results from physics and geometry \cite{BCOV-93}.  Note that
at other genera than $r=1,2$, there appears to be no difference in the
invariants.

\section{String Theory Approach} \label{sec:st-approach}

This section is entirely physical in substance and contains no
computation. It presents the link from GW theory to topological and
type IIA string theories and the meaning of the GW invariants in the
latter context (then denoted $\tilde n^r_d$).  They are degeneracies
of bound states of D0 and D2-branes for given charge and spin of the
BPS state.  Chapter~\ref{sec:relation-het} will complete the physical
picture by describing another context in which GW potentials can be
computed: heterotic strings.

\subsection{Topological Strings} \label{sec:top-strings}

Let us now recall how M-theory arrives at the above results and what
meaning it gives to the variables.  $ N=2 $ SCFT's are the building
blocks for string theories.  In particular, we obtain a topological
theory (A model, resp. B model) by twisting the fermion numbers of the
CFT (K\"ahler twisting, resp. complex twisting). Moreover, after
coupling it to gravity ({\em i.e.}~after insertion of $3g-3 $ Beltrami
operators), the theory allows for a partition function at genus $ g$,
denoted by $F_g$.  In case of the weak coupling limit of the A model,
the latter superpotential term should be a {\em holomorphic} function of the
K\"ahler moduli $t^i $ only, since it involves only chiral superfields
(so-called F-terms).  Hence this limit is obtained by letting the
anti-holomorphic moduli tend to infinity, $ \bar{t} \rightarrow
\infty$, also known as the holomorphic limit or topological limit.
Most interesting cases are susy sigma models with target space a
K\"ahler manifold (which we will take to be a Calabi-Yau $X$, for
convenience): they yield an $N=2 $ QFT whose only finite terms in the
action -- in the above limit -- are those for which $ \bar{\partial}
X_i =0 $, {\em i.e.}~the bosonic coordinates on the target space should be
holomorphic functions.  From the close analogy between an $ N=2 $ $
SCFT$ and the bosonic string theory, we are led to introduce Riemann
surfaces $ \Sigma_g $ and interpret the bosonic coordinates $X_i $ to
describe holomorphic maps from $ \Sigma_g $ to the Calabi-Yau $X $.
Overall, we re-interpret the topological partition function $F_g $ as
counting holomorphic maps, or what amounts to the same, their image,
{\em i.e.}~holomorphic curves embedded in $X $.

\subsection{Link with Type IIA Theory}

The question arises as to which string theory has a similar
description?  From what we already said about the topological
partition function $ F_g$ and also recalling that it is left-right
symmetric as well as related to twisting of a susy sigma model for
closed string theory, we are led to consider amplitudes with $2g-2$ RR
fields of charge $3/2$ on a Riemann surface of genus $g$.
Incidentally, type IIA superstrings compactified on a CY threefold
yields an $N=2$ theory, and $2g-2$ graviphoton vertices just satisfy
the requirements.  Indeed, it can be verified that its low-energy
effective action coupled to supergravity contains the following
Lagrangian F-terms:
\begin{equation}  \label{F-terms}
   \int F_g  R^2_+ F_+^{2g-2}   
\end{equation}
where the insertions of the graviphoton fields $ F_+ $ account for the
change from ordinary type IIA ({\em i.e.}~non-twisted) theory to a
properly twisted theory.  $ R_+$ denotes the self-dual part of the
Riemann curvature of the compactification manifold $X$ ({\em
  i.e.}~insertion of two gravitons).  These F-terms only appear at
genus $g$ (see section~\ref{sec:moduli}) and do not receive other
quantum corrections ($\hbar$ corrections), not even at
non-perturbative level, as the dilaton of type IIA belongs to a
hypermultiplet while N=2 Sugra forbids adependance of the $F_g$ on
matter hypermultiplets.
Therefore the tree-level prepotential $F_0$ (whose K\"ahler parameters
$t_i$ are in vector multiplets) is {\em exact} at the full quantum
level. However, it will suffer worldsheet instanton corrections
($\alpha'$ corrections, or rather $\alpha'/R^2$ corrections where $R$
is the K\"ahler parameter -- or radius -- of the compactification
space), so a neat trick is to compute it via mirror symmetry, where
K\"ahler and complex moduli are swapped.  In any case, topological
string theory offers a much easier way to compute the above F-terms.

Twisted type IIA compactified on the Calabi-Yau $X$ re-interprets
topological amplitudes ({\em i.e.}~counting holomorphic curves) as corrections
to $R^2_+ F_+^{2g-2} $.  That is, $F_g$ is understood as the coupling
of the action terms $R^2_+ F_+^{2g-2} $.  When viewing $X$ as a
$K3$-fibration over the base $\P^1$, the weak coupling limit is
the large volume limit, say large volume of the base.  Since these
amplitudes are exact beyond genus $g$ (free of quantum corrections
corrections), we might just as well compute them in the regime of
strong coupling constant, where M-theory comes out of hiding.  At any
rate, topological string theory computes exact quantities of the
physical type IIA string theory.

\subsection{M-Theory and the Schwinger Computation}
\label{sec:schwinger-one-loop} 

The trick we shall use consists in integrating out light BPS states.
Alas, in the large CY limit, D-branes grow in size and thus in mass.
So we are looking for an additional limit, overriding the first one,
where D0 and D2-branes become the lightest states.  This is where
M-theory enters the game: since M-theory compactified on a circle
$S^1$ of radius $R_{10}$ tends to type IIA as $ R_{10} \to 0 $ and
since $g_s = 2 \pi R_{10}$, strong coupling (large $g_s$) is but the
decompactification limit of IIA where one recovers the full 11d
M-theory.  The advantage of this limit is that the relevant ({\em ie}
lightest) BPS states are then D0 branes and D2 branes of IIA string
theory, {\em i.e.}~in terms of M-theory: KK modes and M2 branes respectively,
or bound states thereof.

When M-theory is additionally compactified on the Calabi-Yau $X$, the
large volume of the latter gives us back the perturbative regime of
type IIA, where the $F_g$ are the topological string partition
functions and are given by the worldsheet instanton sum
(\ref{potential}); that is, the images of the holomorphic maps
$\Sigma_g \to X $ are just the supersymmetric cycles on which wrap the
M2 branes of our M-theory.  Note that the two compactification limits
of large circle and large CY have an opposite effect on the string
coupling: the first increasing it, the second lowering it.  We choose
the limits such that the first effect be the dominant one, {\em i.e.}~strong
coupling, where the lightest states are the D0- and D2-branes.

Note also that our strategy of integrating out light states has close
analogy with Seiberg-Witten theory, where light magnetic monopoles
become massless at singular points of the moduli space.  These
generate monodromies for the scalar Higgs field ($a\sim \< \phi \> $)
and its dual ($\hat a = \p \CF / \p a $), and are integrated out via a
one-loop integral.  This integral is the only field theory
contribution.  There is no tree-level contribution, as SW-theory is a
free theory near a singularity ({\em i.e.}~where dyons or monopoles
become massless).  It is also exact at one-loop, so there is no
question of higher loop contributions.  The same applies to the
Schwinger one-loop calculation below, {\em i.e.}~it captures both
perturbative and non-perturbative parts: the one-loop calculation is
the exact result.

The variables occurring in $\tilde{F}$ of (\ref{F}) have the following
physical meaning: The graviphoton field strength $F_+$ has been
absorbed into the string coupling constant $ g_s $ to form the
parameter $ \lambda = g_s F_+ $ in $ F= \sum_0 \lambda^{2g-2} F_g $.
The quantised momentum of a BPS state around the compactification
circle $S^1$ is labelled by an integer $n$, which is summed over all
the spectrum and turned into the integer $k$ of (\ref{F}) after
Poisson resummation.  The corrections (\ref{F-terms}) are calculated
in \cite{GV1-98} via a one-loop Schwinger integral with BPS states running
around the loop, corresponding to bound states of D2-branes with $n$
D0-branes.  The latter have a charge (=mass) $e=m=|Z|$ with $ Z= (A +
2 \pi i n) /g_s $ where $n \in \Z$ is their momentum quantum
number and $A$ is their area: $A=0$ for D0-branes (these account for
the constant map contribution), or $ A = d \cdot t$ for D2-branes
(these account for the contribution from rational curves in the class
$d$).  These D2-branes are assumed to have the topology of $S^2$, but no
spin.  The one-loop Schwinger integral computes the free energy of a
charged scalar (charge $e$, mass $m$) in a constant self-dual field
$F_+$:
\begin{equation}  \label{schwinger}
  \begin{split}
F(\lambda ) &= \int_\epsilon^{\infty} \frac{ds}{s} \ \textrm{tr }
e^{-s(\Delta +m^2)} \\ &= \int_\epsilon^{\infty}  \frac{ds}{s}
\frac{e^{-sZ/F_+}}{(2 \sinh s/2)^2}
  \end{split}
\end{equation}
which we sum over all momenta $n \in \Z, n\neq 0$, and
Poisson resum via  $\sum_{n \in \Z} e^{inx} = \sum_{k
  \in \Z} \delta (x-2 \pi k) $ to obtain:
$$
F(\lambda) =  \sum_{k\geq 1} \frac{1}{k} \frac{q^{-dk}}{(2 \sinh
  \frac{k g_s F_+}{2})^2}      
$$
This yields (\ref{F}) upon taking all degrees into account weighed
by $ \tilde{n}^r_d$, redefining the topological string coupling
constant $\lambda = g_s F_+ \to i \lambda$, swapping the K\"ahler cone
($t_i \to -t_i$), as well as incorporating the spin of the BPS state;
mysteriously enough, the contributions from higher genus curves are
obtained not by explicitly changing the topology of D2-branes but by
allowing them to have spin:   
$\textrm{tr} \ e^{-2 k \lambda J_3} \sim (2 \sin k \lambda/2)^{2r}$
for spin content $I_1 \otimes I_r$ (see section~\ref{sec:physical-meaning}).
This accounts for the extra sine term in (\ref{F}) and the necessary
$r$-dependence.  Maybe one could invoke the $4d$ gravitino $\psi^\alpha
= \psi^\alpha_\mu dx^\mu $ to account for this $r$-dependence, as
there are $r$ holomorphic one-forms $dx^\mu$ on a curve (=brane) of
genus $r$.

\subsection{Physical Meaning of the $\tilde{n}^r_d$}
\label{sec:physical-meaning} 

The invariants $\tilde{n}^r_d$ already introduced in eqns
(\ref{physical-n^r_d}) and (\ref{F}) have the following
interpretation: The BPS states are equivalent, in the M-theory
perspective, to M2 branes wrapping susy cycles in a given homology
class $d$.  The latter class also describes the mass (or charge, or
tension) of the BPS states -- {\em i.e.}~the area $ d\cdot t$ of the M2 brane
-- in terms of the K\"ahler moduli $ t_i$ of the Calabi-Yau.  An
additional label is the transformation property under the spatial
Lorentz group $ SO(4) = SU(2)_L \times SU(2)_R $ in $4+1$ dimensions,
given by two half-integers $ (j_L, j_R) $.  If we denote by $N^d_{j_L,
  j_R}$ the number of BPS states with these quantum numbers, then the
latter transform in the following representation:
$$ 
\big[ (\half,0) \oplus 2(0,0) \big] \otimes
\bigoplus_{j_L,j_R} N^d_{j_L,j_R}\ [(j_L,j_R)].
$$
The left-moving content, $[(\half)+2(0)] \otimes [j_L]$, is the
usual N=2 BPS multiplet for a spin $j$ ground state.  We shall make
use of the basis $I_r := [(\half) \oplus 2(0)]^{\otimes r} $, or
explicitly:
$$
\begin{array}{rl}
  I_0 &= (0)\\
  I_1 &= 2(0) \oplus (\frac{1}{2}) \\
  I_2 &= 5(0) \oplus 4(\frac{1}{2})  \oplus  (1)\\
  I_3 &= 14(0) \oplus 14(\frac{1}{2})  \oplus 6(1) \oplus (\frac{3}{2})\\
  ...
\end{array}
$$

Since $F_+$ only couples to the left spin quantum numbers, only the
degeneracy of the left spin content will be an invariant of the BPS
spectrum.  That is, we need to sum over the right representation and,
when expressing the result in a suitable basis, we obtain our
sought-for invariants $ n_d^r$ of the theory:
$$ 
\bigoplus_{j_L}
\Big( \sum_{j_R} N^d_{j_L,j_R}(-1)^{2 j_R}(2j_R+1) \Big) [(j_L)]
=: \bigoplus_r \tilde{n}_d^r ~I_r
$$ 
where $ (2j_R+1)$ is the right spin degeneracy, and $ (-1)^{2 j_R}$
accounts for bose/fermi statistics.  This defines the $ \tilde{n}_d^r$ in
terms of the basis $I_r$.  

The merit of the basis $I_r$ is that tr$_{I_r} \dots = ($ tr$_{I_1}
\dots)^r$, so that a trace over the above representation content yields
$$
{\rm tr}_{\sum_r \tilde n^r_d I_r} ~(-1)^{2J_3} y^{2J_3} = \sum_r
\tilde n^r_d ~({\rm tr}_{I_1} ~(-1)^{2J_3} y^{2J_3} )^r = \sum_r
\tilde n^r_d ~(2-y-y^{-1})^r.
$$
We have placed ourselves in the context of N=2 compactifications,
where $y$ keeps track of the third component of the $SU(2)$ current
$J$ ({\em i.e.}~of the $U(1)$ subgroup).  

For instance, if we have four left-moving bosonic oscillators
$\alpha_{-n}^i$ with $SU(2)_L \times SU(2)_R$ (space-time) content
$(\half,\half) $, their partition function is
$$
\prod_{n\geq 1} {1\over (1-yq^n)^2 (1-y^{-1} q^n)^2 } 
= \sum_{d\geq 0} \big( \sum_{r=-d}^d c^r_d ~y^r \big) q^d
= \sum_{d\geq 0} \big( \sum_{r=0}^d \tilde n^r_d ~(2-y-y^{-1})^r \big) q^d
= {\rm tr} ~(-1)^{2J_3} y^{2J_3} q^{L_0}
$$
with the trace again over $\sum_r \tilde n^r_d I_r $.  Concretely, at
level $d=3$, we have $\tilde n^r_3 =40, -60, 28, -4$ for $r=0,1,2,3$.
Of course, this coincides with taking all combinations of oscillators
at level 3, namely $\alpha_{-1}^i \alpha_{-1}^j \alpha_{-1}^k |0\> $,
$\alpha_{-1}^i \alpha_{-2}^j |0\> $, $\alpha_{-3}^i |0\> $, and
summing over the right spin content and weighing each term by
$(-1)^{2j_R} (2j_R+1)$:
$$
\begin{array}{rl}
\big[ ({3\over 2}, {3\over 2}) \oplus (\half,\half) \big] \oplus \big[
(1,1) \oplus (1,0) \oplus (0,1) \oplus (0,0) \big] \oplus \big[
(\half,\half) \big] \longrightarrow & 4(0) -4(\half) + 4(1) -4({3\over
  2}) \\
&= 40 I_0 -60 I_1 +28 I_2 -4 I_3, 
\end{array}
$$
where on the {\em lhs} the three square brackets refer to the content of
the three combinations of oscillators.  Note that for operators of
different level, the tensor product distributes into the brackets:
$(\half,\half)_{-1} \otimes (\half,\half)_{-2} = (\half \otimes
\tilde\half , \half \otimes \tilde\half) = (1\oplus 0, 1\oplus 0)$.

Another instance is a heterotic compactification on $K3 \times T^2$,
where the 24 left oscillators $\alpha_{-n}^i$ with $SU(2)_L \times
SU(2)_R$ content $20 (0,0) \oplus (\half,\half) $ and partition
function
\begin{equation} \label{BPS-het}
\prod_{n\geq 1} {1\over (1-q^n)^{20} (1-yq^n)^2 (1-y^{-1} q^n)^2 } 
= \sum_{d\geq 0} \big( \sum_{r=0}^d \tilde n^r_d ~(2-y-y^{-1})^r \big) q^d,
\end{equation}
which is just the $p^0$ term of equation (6.3) of \cite{KKV-99}; that
is, the $\tilde n^r_d$ have here the enumerative meaning of counting BPS
states wrapped on $K3$. 

By construction, the $ \tilde{n}_d^r$ are integers, as they merely
count BPS states.  That these integer quantities are the same as those
appearing in the Gromov-Witten potential $\tilde{F}$ is a bold claim,
given that our original topological invariants were only expected to
be rational numbers: the counting of holomorphic curves from a Riemann
surface to a threefold will in general involve integrals over moduli
spaces of holomorphic maps, {\em i.e.}~top characteristic classes or Euler
numbers.  So we see that M-theory offers a novel approach to the
Gromov-Witten invariants.

\section{Relation with Heterotic Strings} \label{sec:relation-het}

We now head towards topics related to the GW potential, both physical
and mathematical.  This small section is, much like
section~\ref{sec:st-approach}, devoted to understanding the stringy
background of GW theory and its genus $g$ potential. It is purely
physical, without computations, and presents the one-loop integral of
the heterotic string amplitude and an attempt at solving it via Jacobi
forms.

The M-theory (or type IIA) approach to Gromov-Witten invariants is the
most recent one, but another attempt from string theory proved
fruitful: this is the calculation in heterotic string theory,
following the discovery in 1995 of its description dual to the type
IIA theory.  Given that the perturbative expansion is governed by the
dilaton, that the dilaton lies in a hypermultiplet in type IIA and in
a vector multiplet in heterotic, and that the two kinds of multiplets
do not mix with each other, the duality then allows us to extract
non-perturbative knowledge (instanton corrections,...) in one model
from perturbative expansions in the dual model.  The similar
properties of the topological couplings $F_g $ in heterotic theory
compactified on $K3 \times T^2$ and IIA compactified on a CY threefold
were seen as a test of this $ N=2$ duality.  In particular the $F_g$
satisfy the same holomorphic anomaly equation in the respective weak
coupling limits ({\em i.e.}~$S \to \infty$ and $ t \to \infty$).

\subsection{Moduli and the One-Loop Level} \label{sec:moduli}

In the semi-classical limit ({\em i.e.}~taking the dilaton $S
\rightarrow \infty$), heterotic theory -- with embedding of the spin
connection in the gauge group -- is described by two complex moduli
$(T,U)=:y$, element of the Narain moduli space $\CN^{2,2} =
O(2,2)/O(2) \times O(2) \simeq \CH \times \CH$.  These are the moduli
corresponding to the compactification on $T^2$ and governing its
bosonic partition function.  Since the string coupling equals $g_s =
e^{iS} $, the above limit is the weak coupling limit.  This coincides
with the topological limit on the type IIA side, since one of the
K\"ahler moduli of the Calabi-Yau is given by $S$ itself.  For
instance, in the case of the hypersurface of degree 24 in
$\P(1,1,2,8,12)$ (a $K3$ fibration), the precise map between the
heterotic moduli and the IIA K\"ahler moduli is $t=(U, S, T-U)$, and
sending $S \rightarrow \infty$ means infinite volume of $X$ -- or
rather of the base $\P^1$ of the $K3$-fibration.

An important \underline{difference between the IIA and heterotic}
calculations is that in the former the $F_g$ are generated at $g$-loop
level, while in the latter they all occur at 1-loop level (due to
$N=2$ non-renormalisation theorems)\footnote{Except for $F_0$ and
  $F_1$ which also carry tree-level contributions}.  This can be
argued as follows: the $F_g$'s have the meaning of moduli-dependent
couplings for the low-energy effective action terms $R^2 F_+^{2g-2}$.
They are homogeneous of degree $2-2g$ in the superfields $X^I$, whose
scalar component ($X^0$) has the following dependence on the string
coupling and the K\"ahler potential:
$$ 
F_g(X) = (X^0)^{2-2g} F_g(Z) = \Big(\frac{e^{K/2}}{g_s}\Big)^{2-2g} F_g(Z)
$$
where $Z=X/X^0$ are the moduli.  The K\"ahler potential $K$ depends
on the string coupling $g_s$ via the dilaton: this dependence is nil
in the case of type IIA (as the dilaton is in a hypermultiplet), and
$\log g_s^2$ in the case of heterotic strings.  Accordingly, $F_g(X)$
is of order $g_s^{2g-2}$ or $g_s^0$ respectively in the string coupling.
Counting string loops, this means $g$-loop or one-loop respectively.

In both theories, the prepotential develops logarithmic singularities
reminiscent of the Seiberg-Witten analysis: different branches of the
enhanced symmetry locus (ESL) collapse in the large moduli limit.
On the IIA side, this corresponds to the conifold locus in the moduli
space of Calabi-Yau 3-folds.  On the heterotic side, it corresponds to
codimension 1 surfaces of the moduli space where the gauge group
$U(1)^{n_v+2}$ is enhanced to $SU(2)$ because two vector
multiplets become massless.  Note that the rank of the gauge group is the
number $n_v$ of vector multiplets (containing the compactification
moduli) plus two extra vectors (graviphoton from the sugra multiplet
and the vector in the vector-tensor multiplet of the dilaton).

\subsection{Computing $F_g$: the One-Loop Integral}

As in type IIA, the $F_g$ of heterotic theory on $K3 \times T^2 $ are
couplings for the $R^2 F_+^{2g-2}$ terms, {\em i.e.}~amplitudes involving two
gravitons and $2g-2$ graviphotons.  They are computed using the odd
spin structure on the worldsheet torus with insertion of vertex
operators.  The graviton vertices absorb the space-time fermions,
while the graviphotons contribute $\left( p_R /\sqrt{2T_2U_2}
\right)^{2g-2}$, which will be summed with $q^{\half|p_L|^2}
\bar q^{\half|p_R|^2}$ over the $\Gamma_{2,2}$ lattice of the torus.
The left-moving space-time (transverse) bosons and the extra free
boson for the $U(1)$ current yield $1/\eta^3$, whereas the
conformal blocks of the internal SCFT generate the $K3$
partition function $C_{_{K3}}= \tr_{_{RR}} ~(-1)^{F_R} q^{L_0-c/24}
\bar q^{\bar L_0-\bar c/24}$.  The latter is independent of
$\bar q$, as usual due to Susy for the massive modes and to the
absence of instanton contributions: these would be sensitive to
deformations of hypermultiplet moduli (which contain the K\"ahler
moduli of the compactification space), yet $F_g$ is not sensitive to
them! Overall \cite{AGNT-95}:
\begin{equation*}
  \begin{split}
      F_g = {1\over 2 \pi^2} {1\over (g!)^2}
  \int_\CF {d^2\tau\over\tau_2^3} & \frac{1}{\eta^3} \<
  \prod_{i=1}^g \int_{T^2} d^2 x_i Z^1 \p Z^2 (x_i)
  \prod_{j=1}^g \int_{T^2} d^2 y_j \bar Z^2 \p \bar Z^1 (y_j) \> \\
  & \times C_{_{K3}} \sum_{\Gamma_{2,2}} \left( p_R /\sqrt{2T_2U_2} 
  \right)^{2g-2} q^{\half|p_L|^2} \bar q^{\half|p_R|^2}
  \end{split}
\end{equation*}
where $Z$ are the complex coordinates for the space-time right-moving
bosons.  The correlator for space-time bosons $\< \int_{T^2} Z
\partial Z\> $ can be summed over $g$ with
$\big( {\lambda\over\tau_2} \big)^{2g}$ and ${1\over (g!)^2}$ to
yield the function
$$
  \bigg( \frac{ 2 \pi i  \lambda \eta^3 }{
  \vartheta_1 (\tau, \lambda)} \bigg)^2 {\rm e} ^{-\frac{\pi 
  \lambda^2 }{\tau_2}}
$$
which is modular invariant under $PSL(2,\Z)$, {\em i.e.}~under
$\tau\to{a\tau+b \over c\tau+d}$ and
$\lambda\to{\lambda \over c\tau+d}$.  
The full generating function for the amplitudes at all genera
can then be written as
\begin{equation} \label{agnt}
  F(\lambda, T,U)= \sum_{g\geq 1}\lambda^{2g} F_g =  {1\over
  2\pi^2} \int_{\CF} {d^2\tau\over\tau_2} 
  \frac{C_{_{K3}}}{\eta^3} \sum_{\Gamma_{2,2}} \bigg( \frac{ 2 \pi i
  \lambda \eta^3 }{\vartheta_1 (\tau, \tilde \lambda)} \bigg)^2
  e^{-\frac{\pi \tilde \lambda^2 }{\tau_2}} q^{\half |p_L|^2}
  \bar q^{\half |p_R|^2},
\end{equation}
where $\tilde \lambda := {p_R \tau_2 \over\sqrt{2T_2U_2}} \lambda$.
In \cite{MM-98}, the quantity $ C_{_{K3}}/\eta^3$ takes the
value of $E_4 E_6/\eta^{24}$, in agreement with the $K3$ elliptic
genus for a $\Z_2$ orbifold of heterotic compactification with gauge
group $E_8 \times E_7 \times SU(2)$, as we shall see in
(\ref{new-susy-index-concrete}) or (\ref{integrand-result}).  We note
that the lattice sum is not restricted over $q^{\half |p_L|^2} \bar
q^{\half |p_R|^2}$, but also involves the $p_R $ contained in
${\vartheta_1 (\tau, \tilde \lambda)}$.

Note also that {\em w.r.t.}~$(\tau, \bar\tau)$, $\tau_2$ has weight
$(-1,-1)$, ${d^2\tau \over\tau_2}$ has weight $(-1,-1)$, $E_4
E_6/\eta^{24}$ has weight $(-2,0)$ while the remaining parts of the
integrand have no well-defined weights. It is only upon expanding
$\bigg( \frac{ 2 \pi i \lambda \eta^3 }{\vartheta_1 (\tau, \tilde
  \lambda)} \bigg)^2 e^{-\frac{\pi \tilde \lambda^2 }{\tau_2}} $ in
even powers of $\lambda$, say $\lambda^{2k+2}$, and keeping 
$\bar p_R^{2k}$ for the lattice sum, that we obtain a suitable
generalised theta function of weight $(1,1+2k)$: $\Theta(T,U) :=
\sum_{\Gamma_{2,2}} \bar p_R^{2k} ~q^{\half |p_L|^2} \bar
q^{\half |p_R|^2}$.

\subsubsection{Lattice Reduction Technique}

The evaluation of the above integral (\ref{agnt}) over the fundamental
domain is a {\em tour de force} \cite{MM-98}.  It is based on
Borcherds' recursion formula for such automorphic integrals or {\em
  theta transforms}: at each step the lattice for the theta function
is reduced by two dimensions.  Although the results for $F_g$ look
rather messy, their holomorphic limit ($ \bar t \to \infty$) is quite
simple, and one does indeed recover the constant map contribution
(\ref{constant-map}) which had also been obtained by computations on
the type IIA side \cite{BCOV-94}.  The result for non-constant maps in
the holomorphic limit $ \overline{T}, \overline{U} \to \infty $ ($T,U$
fixed) is similar to the $ \tilde{F}_g$ in (\ref{F_g-expl}):
$$
 \tilde{F}_g \sim \sum_{r>0} c_{g-1}(r^2/2) \ Li_{3-2g}(e ^{2 \pi i
   r \cdot y} ) 
$$ 
where $y=(T,U)$ is the heterotic parameter, $ r=(n,m)$ is a point in
the lattice $ \Gamma_{1,1}$, $ r \cdot y = nU + mT $, $r^2=2nm$ and
the condition $ r>0$ stands for $ n,m \geq 0$ or $(n,m)=(1,-1)$ but not
$(0,0) $.  The $ c_{g-1}(n)$ are the Fourier coefficients of the
modular function occurring in (\ref{agnt}):
$$ 
\frac{C_{_{K3}}}{\eta^3} ~G = \sum_{\substack{g \geq 0 \\ m \geq -1}}
c_{g-1}(m) \ \lambda^{2g-2} q^m \ ,
$$ 
$q:=e^{2 \pi i \tau}$.  The computation suggests that the result
should depend on the region of moduli space, and that a wall-crossing
formula will relate different regions.  Here, the wall happens to be
the codimension one surface $ T_2=U_2 $.  By chance, the wall-crossing
behaviour vanishes in the holomorphic limit, as expected from the type
IIA side where this behaviour reflects the fact that two CY threefolds
related by flop transition are birationally equivalent ({\em i.e.}~same Hodge
numbers).

Moreover, by organising the terms in the same way as the type IIA
result, \cite{MM-98} were able to extract from $\tilde{F}_2$ the number of
genus $2$ curves on a particular CY threefold.  To this end, they first
identified $ -2 \sum_{r>0} c(r^2/2) \ \Li_{-1} (e^{2 \pi i r \cdot y})
$ as the term corresponding to rational curves, with $c(n)$ the
Fourier coefficients of $ E_4 E_6 / \eta^{24}$ -- these had already
been seen in a previous heterotic approach \cite{HM-95} (see below) as
counting the number of rational curves.  Since the $c(n) $ vanish for
$ n<-1$, the condition $ r>0$ above can be replaced by the positive
root condition of the $ \Gamma_{1,1}$ lattice for the monster Lie
algebra, {\em i.e.}~the condition $n>0;$ or $n=0, m>0$.  The latter meaning
was adopted in \cite{HM-95}.  

\subsubsection{Attempt Using Properties of Jacobi Forms}

Integrals of elliptic genera over the fundamental domain have also
been tackled from a different perspective in \cite{N-98}.  Rather than
integrating the usual elliptic genus, the latter author slightly
altered that index to obtain the so-called {\em new supersymmetric
  index}, and studied its integral.  He showed that this new index,
which is a trace over the left- and right-moving Ramond sectors,
conveniently factorises when the compactification space is a product
$T^2\times K$ for some two-fold $K$:
\begin{equation} \label{susy-index}
  \textrm{tr}_{_{R\times R}} \ (-1)^F J_0 \bar{J}_0 \ 
  q^{L_0-\frac{\hat{c}}{8}} \bar{q}^{\bar{L}_0-\frac{\hat{c}}{8}} =
  \left( \sum_{(p_L,p_R)\in\Gamma_{2,2}} q^{\frac{1}{2}p_L^2}
    \bar{q}^{\frac{1}{2}p_R^2} \right) \Big( \textrm{tr}_{K} \ (-1)^F
    q^{L_0-\frac{\hat{c}}{8}} \bar{q}^{\bar{L}_0-\frac{\hat{c}}{8}}
    \Big),
\end{equation}
where $\hat{c}$ is the dimension of the target space (3 in our case).
The bracketed sum is the famous bosonic partition function for the
torus, coming from the trace over the torus part of the threefold
$K\times T^2$.  The second part comes from the two-fold $K$ and has
to do with its elliptic genus at $z=0$, $\Phi(\tau,0)$. 

The full elliptic genus reads $\Phi(\tau,z)= \textrm{tr}_{_{R\times
    R}} \ (-1)^F q^{L_0-\frac{\hat{c}}{8}} y^{J_0}$, see also
(\ref{ell-genus}), and is a weak Jacobi form of weight 0 and index $
\hat{c}/2$ (see appendix \ref{sec:jacobi-forms}).  For twofolds with
$SU_2$ holonomy (two-tori, $K3$ surfaces), its expression is
well-known \cite{KYY-93}; for $ \hat{c}=1,2,3$ theories, $
\Phi(\tau,0)$ boils down to the {\em Witten index} of the theory, {\em
  i.e.}~to tr $(-1)^F=\chi(K)$, the Euler character of the target
space.  For general models, one recovers the Euler character only upon
taking the limit $ \lim_{\tau \rightarrow i\infty} \Phi(\tau,0) $.

However, in \cite{N-98}, $\Phi(\tau,0)$ was not merely the Euler character,
because a Wilson line was introduced.  This inserted an extra
compactification modulus (next to $T$ and $U$ for the torus) which
prevented from writing the CY threefold as a direct product $ K\times
T^2$.  Thus, the decomposition property (\ref{decomposition}) of Jacobi forms
was used to compute the integral.  The property states that a Jacobi
form can be written as a finite linear combination of theta functions
with modular forms as coefficients.  For $z=0$ the claim reduces to
$$
\Phi(\tau,0)= \sum_{\mu (\textrm{mod } 2m)} h_\mu(\tau)
\sum_{r\equiv \mu (\textrm{mod } 2m)}  q^{r^2/4m},
$$ 
and $m=\frac{\hat{c}}{2}=1$. The crux is that the extra summation
of $q^{r^2/4m}$ over $r$ could be joined with $q^{\frac{1}{2}p_L^2}$
in the above (\ref{susy-index}), resulting in an overall sum over a
$(3,2)$ lattice rather than just over $\Gamma_{2,2} $. In other words,
one simply ended up with a new Siegel theta function of signature
$(3,2)$, something one could integrate over the fundamental domain --
using the well-known technique of `unfolding'.

Do we stand a chance for a similar trick for the integral in
(\ref{agnt})~? Indeed, we recognise $-\eta^6/\vartheta_1(\tau,z)^2$ to
be the inverse of the weak Jacobi form $A_{-2,1}$ studied in appendix
A.4 of \cite{G-04}.  Unlike for the elliptic genus above, $z$ is not
set to 0 but contains factors of $\lambda$ (string coupling) and $
\tau_2$.  But this should not discourage us, since these $ \tau_2$
factors may be joined with the $\tau_2$ factors $ q^{\frac{1}{2}p_L^2}
\bar{q}^{\frac{1}{2} p_R^2} = e^{\pi i \tau p^2 - 2\pi \tau_2 p_R^2} $
occurring in the Siegel theta function.

Yet a more serious point flaws our attempt to solve the integral by
changing the Siegel theta function into a theta function of signature
$(3,2)$, namely the fact that inverses of Jacobi forms have negative
index and hence do not enjoy the periodicity property
(\ref{periodicity}) nor the decomposition into linear combinations of
theta functions (\ref{decomposition}).  This too is explained in
appendix A.4 of \cite{G-04}.  So the trick of \cite{N-98} cannot be
used here.

\section{Automorphic Properties}\label{sec:automorphic-prop}

This section deals with the mathematical aspects of the full GW
potential in product form, and is essentially mathematical -- though
the first three sections present results without derivations. The
evaluation of an integral over the fundamental plane \cite{HM-95}
yields the logarithm of a infinite product, of which Borcherds
\cite{B-95} had already predicted the automorphic properties (though
obvious in this approach). The generalisation to arbitrary GW
potentials is tried with Borcherds' lifting of Jacobi forms to
automorphic forms, but remains inconclusive.  Prerequisites are
section~\ref{sec:gw-results-CY} and familiarity with infinite products
\`a-la Borcherds; results will not be used in later sections.

\subsection{Torus as Target Space}

If we replace our three-dimensional Calabi-Yau space by a
one-dimensional one, {\em i.e.}~by a mundane elliptic curve, then the
Gromov-Witten problem boils down to the Hurwitz problem of counting
covers of a Riemann surface.  The {\em free energies} $F_g$ have been
similarly defined, and an explicit expression for the {\em partition
  function} $Z:=\exp \sum_1 \lambda^{2g-2} F_g $ was given in
\cite{Dou-93}.  It involves a generalised theta function, thereby ensuring
modular properties of the $F_g$'s.  For example, $F_1 = -\log \eta(q)$, 
and $F_2$ is a linear combination of Eisenstein series \cite{Ru-94}.  This
is to be compared with the GW potential $F_1 = -\sum_d n^1_d \log
\eta(q^d)$ where we have an additional sum over the homology class of
the image curve.  Of course, in the case of covers of an elliptic
curve, there is no degree to keep track of, and $t$ is the K\"ahler
modulus of the flat torus.  Mirror symmetry relates $t$ to the complex
modulus $\tau $ of the mirror elliptic curve, which accounts for its
modular covariance under $ PSL(2,\Z)$.  More generally, $F_g$ is a
quasi-modular form of weight $6g-6 $.  One might then wonder whether
our present GW potentials enjoy similar modular properties.

However, the occurrence of the $\tilde{n}^r_d$ in (\ref{F_g-expl}) and the fact
that $d$ is a tuple (and not just an integer) spoil the following
naive hope of modular properties:
$$ 
F_g \sim \sum_{d>0} \Li_{3-2g}(q^d) = \sum_n n^{2g-3}
\frac{q^n}{1-q^n} = E_{2g-2}
$$ yielding the Eisenstein series of weight $2g-2$.  So one wonders if
a favourable choice for the integers $\tilde{n}^r_d$ and for the summation
over $d$ would keep the modular properties.  This was indeed the case
for part of the results of \cite{HM-95}, to which we now turn.

\subsection{Harvey-Moore and the Theta Transforms}

In that work, the prepotential for heterotic string theory, which
coincides with $F_0$, was computed by hand.  By equating the Wilsonian
coupling with the one-loop coupling renormalisation
(\ref{one-loop-IR-finite}), one obtains a differential equation for
the prepotential, involving a class of integrals over the fundamental
domain.  The integrand can be replaced by the ``new susy index'', as
in (\ref{one-loop-result-bis}), which can be explicitly computed and
yields a genarlised theta function (or lattice function
$\Gamma_{n+2,2}$) times a modular function, see for instance
(\ref{threshold-E_8}) or (\ref{threshold-E_7}).  Thus these are
integrals of the form of a {\em theta transform}, {\em i.e.}~of an
integral over the fundamental domain of a Siegel theta function times
an almost holomorphic modular function of weight $-s/2$ with Fourier
expansion $F(q)=\sum c(n,k) \ q^n \tau_2^{-k}$, $q=e^{2 \pi i \tau}$,
with summation running over $ n \geq -n_0 $ and $ k=0,1,\ldots,k_0$
for some non-negative integers $n_0$ and $ k_0$:
$$
\Phi_{s+2,2} (y) := \int_{\CF} \frac{d^2 \tau}{\tau_2} \ \Theta(\tau,y) \ F(\tau) \ .
$$ 
This is roughly the Howe correspondence, sending automorphic
functions $F$ for $SL(2,\Z)$ to automorphic functions $\Phi$ for
$O_{s+2,2}(\Z)$.  In general, $F$ is allowed to be modular covariant
at level $N$ (or, equivalently, vector-valued at level 1) and up to a
character, to have poles at cusps and even to have rational powers of
$q$ in its expansion.  For instance, the function $ 1/\tau_2$ is
modular of weight $(1,1)$.  The theta function of weight $(s/2 +1,1)$
is defined for an even self-dual lattice $\Gamma_{s+2,2}$, (with
$8|s$):
$$ 
\Theta (\tau,y) = \sum_{p \in \Gamma_{s+2,2}} q^{p_L^2/2} \
\bar{q}^{p_R^2/2}
$$
where $y$ lies in the Grassmanian $ G(s+2,2) = O(s+2,2)/O(s+2)
\otimes O(2) $ sometimes also called the generalised upper half plane
$ \mathcal{H}^{s+1,1} \cong \R^{s+1,1} + i \ C_+^{s+1,1} $ for the
positive light cone $C_+^{s+1,1}$.  Then $y$ corresponds to a choice
of a positive definite $ s+2$ dimensional subspace of $\R^{s+2,2}
\cong \Gamma_{s+2,2} \otimes \R $, so that every lattice vector $p$
can be projected onto left and right (or positive and negative
definite) subspaces: $ p=(p_L,p_R) $ with $p^2= p_L^2 - p_R^2 $ and $
p_{L,R}^2 \geq 0$.  Note that even though $y \in \mathcal{H}$, the
function $\Theta$ -- and hence $\Phi$ -- is actually automorphic under
Aut$(\Gamma_{s+2,2}) = O_{s+2,2}(\Z) $.  In a more general treatment
\cite{B-96}, $\Theta$ also depends on a homogeneous polynomial and its
defining lattice need not be self-dual (in which case the modular
properties are recovered by considering vector-valued
theta functions).

\subsection{The Result for the Prepotential} \label{sec:result-prepot}

In the easier case where $F(q)=:\sum_{n\geq -n_0} c(n) ~q^n$ is a
meromorphic modular form, \cite{HM-95} have computed the integral by a
generalisation of the technique of \cite{DKL-91} via unfolding the
fundamental domain.  It yields the logarithm of an automorphic product
\`a la Borcherds for the lattice $ \Gamma_{s+1,1} $:
$$
\Phi_{s+2,2}(y)
=- 2 \log \bigg|
 e^{- 2 \pi \rho\cdot y}
\prod_{r>0 } \big( 1-e^{- 2 \pi  r \cdot y  } \big)^{c(-r^2/2)}
  \bigg|^2
+ c(0) \big( - \log [-(\Re y)^2] -\CK \big),
$$
with $r$ the positive roots of the lattice $\Gamma_{s+1,1}$, $\rho$
the so-called Weyl vector of the lattice, and $\CK$ some insignificant
constant.

According to Borcherds\footnote{Note our exponent of $-2\pi r\cdot y$
  as opposed to Borcherds' $2\pi i r\cdot y$.} \cite{B-95}, if $F(q)$
has weight $-s/2$ and the $c(n)$ are integers (with $24|c(0)$ if
$s=0$), then such a product can be analytically continued to a
meromorphic automorphic form of weight $c(0)/2$ on $O_{s+2,2}(\Z)$;
its zeroes and poles lie on rational quadratic divisors $a y^2 +
r\cdot y + c=0$ ($a,c \in \Z$).  Note that inside the radius of
convergence, there are no poles and all zeroes lie on linear divisors
$r\cdot y + c=0$ ($c \in \Z$).

A second type of integral, $\tilde \Phi_{s+2,2}(y)$, in which $F(q)$
is replaced by $F(q)(E_2(q)-{3\over \pi\tau})$, was also computed in
\cite{HM-95} and yielded a similar result containing the above $\log
\prod_{r>0} (1-e^{-2\pi r\cdot y} )^{c(-r^2/2)} = \sum_{r>0}
c(-{r^2\over 2}) \log (1-e^{-2\pi r\cdot y}) = -\sum_{r>0}
c(-{r^2\over 2}) ~{\rm Li}_1(e^{-2\pi r\cdot y}) $ together with
further polylogs Li$_2$ and Li$_3$.

Hence for the linear differential equation where these two integrals
occur, we also expect its solution to share the automorphic
properties.  This solution is the one-loop prepotential; explicitly:
\begin{equation}  \label{hm}
F_0(y) := h^{(1)} (y) =  {1 \over  384 \pi^2} \tilde d_{ijk} y^i y^j y^k
- {1\over 2 (2 \pi)^4} c(0) \zeta(3)
-  { 1 \over  (2 \pi)^4 }
 \sum_{r>0} \textstyle c(-{r^2\over 2}) \ \Li_3 (e ^{-2\pi r \cdot y}) 
\end{equation}
Here $y=(\vec{y},T,U) \in
\mathcal{H}^{s+1,1}$ and $y^2 =\vec{y}^2 + 2TU $.  The symmetric
tensor $\tilde d_{ijk}$ depends on the Weyl vector and the particular
algebra at hand; it is irrelevant for us.  This prepotential was
obtained in the context of heterotic compactifications on $K3 \times
T^2$, with standard embedding yielding a gauge group $E_7 \times SU(2)
\times E_8 \times U(1)^4$.  The $E_8$ part can be broken by the
introduction of $s$ Wilson lines (same $s$ as in $\CH^{s+1,1}$).
\cite{HM-95} considered the cases $s=0$ and 8.

For the \underline{case $s=0$}, $y=(T,U)$, $c(n)$ are the Fourier
coefficients of $ F_{s=0} := E_4 E_6 /\eta^{24} $ (which is $\Gamma_8
=E_4$ times $F_{s=8}$) and the sum runs over all positive roots
$r=(n,m)$ of the monster Lie algebra: $m>0;$ or $m=0, n>0$.  The
Yukawa coupling $ \partial_U^3 h^{(1)} $ agrees with
another expression \cite{AFGNT-95}:
$$
\p_U^3 F_0(T,U) = -{1\over 2 \pi}
\left(
  1-\sum_{r>0} c(kl) ~l^3~ {e^{-2\pi(kT+lU)} \over 1-e^{-2\pi(kT+lU)} }
\right)
= -{1\over 2 \pi} {E_4(iU) E_4(iT) E_6(iT) \over (J(iT) -J(iU) \eta(iT)^{24} )}
$$
This involves only modular forms in $T,U$ separately and is of
weight $(-2,4) $ in $ (T,U)$.  Note that the $SL(2,\Z) \times SL(2,\Z)
$ symmetry is isomorphic to the symmetry group $ SO(2,2,\Z) / \Z_2 $,
where $\Z_2 $ stands for the exchange of $T$ and $U$, and $ SO(2,2,\Z)
$ is the automorphic group of $ \Gamma_{2,2} $.  How much of these
modular properties are then left for the prepotential $ h^{(1)}$
itself~?

For the \underline{case $s=8$}, to which we turn for the remainder of
this section, $c(n)$ are the Fourier coefficients of $ F_{s=8} := E_6
/\eta^{24} $ -- see eqn (\ref{integrand-result}) -- and the sum
runs over the positive roots $ r=(\vec{r},n,m) $ of the $ E_{10} $ Lie
algebra where $\vec{r}$ is itself a positive element of $\Gamma_8$
(the root lattice of $E_8$).

In the particular $\Z_2$ orbifold limit of the $K3$, the massless
spectrum consists of ({\bf 56,2,1}) + 8 ({\bf 56,1,1}) + 32 ({\bf
  1,2,1}) + 4 ({\bf 1,1,1}) in the gauge group $E_7 \times SU(2)
\times E_8 \times U(1)^4$, as argued in the paragraph containing
(\ref{right-moving-fermions}).  These total of 628 hypermultiplets and
388 vector multiplets (388 = rank of the gauge group) are too high to
hope for an easily describable CY dual on which to compactify the type
II A theory.  For the latter, we need few K\"ahler moduli, {\em
  i.e.}~few vector multiplets on the heterotic side, {\em i.e.}~a
smaller gauge group.  As announced, we assume that $E_8$ is broken to
$U(1)^8$ by introducing $s=8$ Wilson loops; we shall further Higgs
completely $E_7 \times SU(2)$ to end up with the gauge group
$U(1)^{12}$.  The Higgsing will cost us 133 + 3 scalars to give mass
to the gauge fields in the adjoint (as outlined in the examples of
section 7 of \cite{G-04}).  Thus we are left with 492 hypers and 12
vectors, which begs for a CY threefold with $h^{21}=491 $ and $
h^{11}=11$.

Such a CY $X$ fortunately exists; it is a degree 84 hypersurface in
$\P(1,1,12,28,42)$, {\em i.e.}~a $K3$ fibration over $\P^1$ where the
$K3$ is a degree 42 hypersurface in $\P(1,6,14,21)$.  In this model for
topological string theory, we see that the above prepotential $F_0$
has remarkably the shape desired for counting instantons -- recall the
Li$_3$ from (\ref{low-genus-F_g}).  Then we see that the rational
holomorphic curves in the fibres of the CY are parametrised by the
positive roots $r$ of the $E_{10}$ root lattice, and their number is
given by $c(-r^2/2)$.  So the only question left is whether $ F_0$ (or
any derivative thereof) in this case enjoys similar modular properties
as in the case $s=0$.

Furthermore, one is driven to ponder on the following issues: What is
the relation between the sum over $d>0$ in (\ref{low-genus-F_g}) and the
sum over $r>0$ in (\ref{hm})?  Would the $F_g$ ($g>0$) in
(\ref{F_g-expl}) enjoy modular or automorphic properties?  Could they
also be expressed as a sum over $ \Gamma_{s+2,2}$ or over a root
lattice of some algebra, rather than over $H_2(X,\Z) = \Z^{h^{11}}$~?
If so, what is the relation between the CY $X$ and the lattice of
which the positive roots govern the counting of holomorphic curves?
Would the $\tilde{n}^r_d$ be the (integer) coefficients of some
nearly-holomorphic modular forms, like in (\ref{hm})?

We conjecture that this should indeed be the case.  The work of
\cite{KY-00} sheds some light in this direction.

\subsection{Extension of the Moduli Space}

The way to convert the general form of (\ref{F_g-expl}) into a modular
product \`a la Borcherds is to generalise Borcherds' lifting of a
Jacobi form $\Phi_0 (\tau,z)$ of weight zero for a positive definite
lattice to a lifting of a form $\Phi_0 (\tau,z,\lambda)$ defined on a
{\em Lorentzian} lattice $\Gamma$.  This extension comes along with the
extension of the moduli space to include the string coupling constant
$\lambda $ next to the moduli $t_i$: $H^2(X,\Z) \oplus H^0(X,\Z) $.
The crucial idea in \cite{KY-00} is to rewrite the GW potential as a sum
over Hecke operators $V_l$ acting on $\Phi_0$:
$$
F = \sum_{g \geq 0} \lambda^{2g-2} F_g (q_i) =  \sum_{g \geq 0}
\lambda^{2g-2} \mathcal{F}_g (p,\tau,z) =  \sum_{l \geq 0} p^l \ \Phi_0
\vert_{V_l} (\tau,z,\lambda) \ ,
$$ 
{\em i.e.}~to absorb the parameter $\lambda$ into the lattice and to
express $F_g$ using variables $(p, \tau,z)$ instead of the tuple
$q_i=e^{t_i}$.  The new lattice is defined to be $ \Gamma := Q^\vee(m)
\oplus \< -2 \> $ where $Q^\vee$ is the coroot lattice of a simple Lie
algebra $ \mathfrak{g}$ and $ Q^\vee(m) = (Q^\vee, m \< \ ,\ \> )$.
The lattice $ \Gamma^*$ contains points $(\gamma, j)$ which will be
multiplied with the variables $(z,\lambda) \in \Gamma_\C $ to yield $e^{2
  \pi i (\gamma \cdot z + j\lambda )} =: \zeta^\gamma y^j $ in the
Fourier series to follow.  Let also $q:=e^{2 \pi i \tau}$.

To further understand the procedure, we give here concrete formulae:
Let $\phi_{-2,m}$ be a nearly-holomorphic Jacobi form of weight $-2$
and index $m$, invariant under the action of the Weyl group of
$\mathfrak{g}$.  The function $ ( \vartheta_1(\tau,\lambda) /
\eta(\tau)^3)^2 $ is itself a weak Jacobi form of weight $-2$ and
index 1.  We define $\Phi_0$ and its Fourier and Taylor coefficients
as follows:
\begin{eqnarray*}
\Phi_0 (\tau,z,\lambda) &:=& - \phi_{-2,m}(\tau,z)
   \frac{\eta(\tau)^6}{\vartheta_1(\tau,\lambda)^2} \\ 
   &=:& \sum_{\substack{n \geq -n_0 \\ (\gamma,j) \in \Gamma^*}} D(n,\gamma,j) \
  q^n\zeta^\gamma y^j  \\ 
   &=:& -\sum_{g=0}^\infty \lambda^{2g-2}\varphi_{2g-2,m}(\tau,z)
\end{eqnarray*}
The $ \varphi_{2g-2,m}(\tau,z) $ are quasi Jacobi forms of weight
$2g-2$ since the $\lambda$-expansion of $1/\vartheta_1^2 $ has
(quasi) modular forms as coefficients.

Then one can check that 
$$
\Phi_0\vert_{ V_l}(\tau,z,\lambda) = - \sum_{g \geq 0}
  \lambda^{2g-2} \varphi_{2g-2,m} \vert_{ V_l}(\tau,z)\,,\quad
  (\forall \ l \geq 0)\,
$$
which allows us to express $\CF_g$ directly in terms of the
$\varphi_{2g-2,m}$:
\begin{eqnarray}\label{mathcal-F_g}
 \mathcal{F}_g(p,\tau,z) &=& \sum_{l \geq 0} p^l \ \varphi_{2g-2,m}
 \vert_{V_l}(\tau,z)  \nonumber \\
   &=& \ldots  \nonumber \\
   &=& \frac{c_g(0,0)}{2}\zeta(3-2g) + \sum_{(l,n,\gamma)>0} 
  c_g(l n,\gamma) \ \Li_{3-2g}(p^l q^n\zeta^\gamma)\,,
\end{eqnarray}
where $(l,n,\gamma)>0$ means $l>0$ or $l=0, \ n>0$ or $l=n=0,\ \zeta >0 $,
and the $ c_g $ are the Fourier coefficients of
$\varphi_{2g-2,m}$:
$$
\varphi_{2g-2,m}(\tau,z) =: \sum_{n,\gamma}c_{g}(n,\gamma) \ q^n\zeta^\gamma\,.
$$
A similar action of the Hecke operators is valid on any Jacobi form
$\Phi_k$ of weight $k$:
$$
 \sum_{l=0}^\infty p^l \ \Phi_k\vert_{V_l}
  (\tau,z,\lambda) = \sum_{(l,n,\gamma,j)>0} D(ln,\gamma,j) \
  \Li_{1-k}(p^l q^n \zeta^{\gamma} y^j)\,,
$$
modulo constant terms.
The benefit of having chosen $\Phi_0$ of weight 0 is that the $ \Li_1$
is but a logarithm, and the {\em partition function} can thus be expressed
as an infinite product:
$$
\mathcal{Z}(\sigma,\tau,z,\lambda) = e^\mathcal{F} 
= \exp \sum_{g \geq 0} \lambda^{2g-2} \mathcal{F}_g \ \ \sim  
\prod_{\substack{(l,n,\gamma,j)>0  \\ (\gamma,j) \in \Gamma^*}} (1 - p^l
q^n \zeta^\gamma y^j)^{D(l n,\gamma,j)} \,
$$
modulo the Weyl vector.  Recall that $p=e^{2\pi i\sigma}$, $q=e^{2\pi
  i\tau}$, $\zeta=e^{2\pi iz}$, $y=e^{2\pi i\lambda}$. 

\subsection{Mapping the Moduli $q_i$ to $(u,p,q,\zeta)$}

The question arises as to how do we map the tuple $q_i=e^{t_i}$ to the
variables $(p,\tau,z)$.  We now carry this out for the special case of
a CY threefold $X$ that can be realised both as $K3$ fibration over
$\P^1$ and as an elliptic fibration over a surface $W2$.  We
choose our simple Lie algebra $\mathfrak{g}$ with coroot lattice
$Q^\vee $ to be of rank $s= h^{1,1}(X) -3 $ and such that the Piccard
lattice of a general $K3$ fibre is isomorphic to $ \Gamma_{1,1} \oplus
Q^\vee(-m)$ for the even self-dual Lorentzian lattice $\Gamma_{1,1}$
of signature $(1,1)$.  In other words, we need three more variables
next to the lattice variables $\zeta^{\gamma}$ to correspond to the
K\"ahler moduli $t_1, \ldots, t_{h^{1,1}}$ of $X$.  Let those three be
$u,p,q$, where $u:=e^{t_1}$ is new here: it is related to $q_1=e^{t_1}
$ and will become redundant in the limit of large base $\P^1$
({\em i.e.}~$ t_1 \to \infty$).  This is why $u$ does not appear in the
argument of $\CF_g = \CF_g (p,\tau,z) $.  We furthermore assume that $
\eta(\tau)^{24} \phi_{-2,m}(\tau, z) $ is a Jacobi function of weight
10 and index $m$, equal to $ -2 E_4 E_6$ at $z=0$, and such that
$c_0(0,0) = -\chi(X)$.

The precise mapping between the $t_i$'s and $(u,p,q,\zeta) $ was
suggested by \cite{KY-00} as 
\begin{equation*}
  \begin{split}
    t_1&=\log u-\log q \,,\\ 
    t_2&=\log p-\log q, \\
    t_3&=\log q-(\gamma_0,\log \zeta)\,,\\
    t_{i+3}&=(\Lambda_i,\log\zeta)\,,\quad (i=1,\ldots,s)\,,
  \end{split}
\end{equation*}
for $\gamma_0$  some positive weight and $\Lambda_i$
$(i=1,\ldots,s)$ the fundamental weights of $\mathfrak{g}$.
Thus
\begin{equation*}
  \begin{split}
    q^t &= e^{t \cdot d} = e^{t_1 d_1} e^{t_2 d_2}  e^{t_3 d_3}
    e^{\sum_1^s  t_{3+i} d_{3+i} } \\
    &= \big( u^{d_1} q^{-d_1} \big) \big( p^{d_2} q^{-d_2} \big)
    \big( q^{d_3} \zeta^{-\gamma_0 d_3} \big)  \big(
    \zeta^{\sum_1^s \Lambda_i d_{3+i} } \big) \\
    &= u^{d_1}  p^{d_2} q^{-d_1 -d_2 +d_3} \zeta^{\gamma_0 d_3 + \sum
    \Lambda_i d_{3+i} } \,,
  \end{split}
\end{equation*}
and as $u \to 0$, only $d_1=0$ contributes, and thus the
identification is:
$$
 p^{d_2} q^{d_3 - d_2} \zeta^{\gamma_0 d_3 + \sum \Lambda_i d_{3+i} } 
\cong p^l q^n \zeta^\gamma \,.
$$
When summing over $ d>0$, {\em i.e.}~over all $d_2, \ldots , d_l$ not all
zero, we see that the following three cases occur:
\begin{equation*}
  \begin{split}
l>0, \textrm{ and } n,\gamma \textrm{ arbitrary}, \\
l=0, \, n>0,\, \gamma \textrm{ arbitrary}, \\
l=n=0, \, \, \gamma >0.
  \end{split}
\end{equation*}
These are just the conditions $(l,n,\gamma)>0$ that we needed, as in
(\ref{mathcal-F_g}).  In other words, we have attained our goal of
rewriting $F_g$ in (\ref{F_g-expl}) as a sum over positive roots of a
lattice $ \Gamma_{1,1} \oplus (Q^\vee)^*(\frac{1}{m}) $.  At least,
this is what \cite{KY-00} conjectured:
$$
F_g = \mathcal{F}_g \quad (g \geq 2).
$$
(plus extra constant terms at $g=0,1$: $F_0^{const}, F_1^{const}$).  
This translates into their main conjecture for the partition function:
\begin{equation*}
  Z(q_i,\lambda) = \CZ(\sigma, \tau, z, \lambda) =
  \exp\left(x^{-2}F_0^{const} + F_1^{const}\right) \left[
  \prod_{(l,n,\gamma,j)>0} (1-p^l q^n\zeta^\gamma y^j)^{D(l
    n,\gamma,j)} \right]\,,
\end{equation*}
with $ (l,n) \in \Gamma_{1,1}, \ (\gamma,j) \in \Gamma^* $. This is a
modular product \`a la Borcherds for the lattice $ \Gamma_{1,1} \oplus
\Gamma^* $, with the only difference that $\Gamma$ here is a Lorentzian
lattice instead of a positive definite one.  The term in the
exponential is then the corresponding Weyl vector.  The vector $
v=(\sigma, \tau, z, \lambda)$, such that $ e^{2 \pi i v}:= p q \zeta y
$, would then be element of the Grassmanian corresponding to the
modular product.  Would this be $ O(s+2,3) $~?

\section{Threshold Corrections for Heterotic Orbifolds}

This crowning section is rather intricate, both mathematically and
physically.  It aims at bringing together the different objects
touched upon in the two previous sections, as well as giving the
details for some computations that we left out.  Thus, we shall come
back on the ``new susy index'' (\ref{susy-index}), explain our
explicit values for the $K3$ elliptic genus $C_{K3} /\eta^3 $ in
(\ref{agnt}) and for the functions $F(q)= \sum c(n) q^n$ in section 
\ref{sec:result-prepot}.  In fact, the latter are nothing but
results that pop up in computations of {\em threshold corrections}.

These are upper-half-plane integrals and are presented in
(\ref{one-loop-IR-finite}) of the first subsection, including a gauge
theory factor which we shall neglect until the very last subsection.

An explicit expression, $\Gamma_{10,2} E_6 / \eta^{24}$, for the
integrand of the threshold correction (\ref{one-loop-IR-finite}) is
given in (\ref{integrand-result}), after a direct but cumbersome
calculation via the orbifold partition functions of
section~\ref{sec:het-orb-part-fct} for $ K3 \times T^2 $ (we assume
familiarity with partition functions for bosons/fermions on several
compactification spaces).  The same expression can be obtained after
first re-writing the original integrand (\ref{one-loop-IR-finite}) as the
``new susy index'' (\ref{one-loop-result-bis}) and then evaluating the
latter in the particular case of $ K3 \times T^2 $ to obtain
(\ref{new-susy-index-concrete}).

The last subsection~\ref{sec:examples-threshold-corr} finally
incorporates the gauge theory factor of the integrand of the threshold
correction. The results there are just the functions $F(q)$ or $F(q)
\hat E_2$ used in the integrands of section~\ref{sec:result-prepot}
that yielded the infinite products \`a-la Borcherds.  Similarly, the
$K3$ elliptic genus $C_{K3} /\eta^3 $ of (\ref{agnt}) will be seen
here to yield $ E_4 E_6 / \eta^{24} $ as in (\ref{integrand-result}).
Thus this chapter connects a good part of what we introduced
previously.

\subsection{Effective Field Theory and One-Loop Threshold Corrections}

Threshold corrections to coupling constants are the differences
between calculations in two separate frameworks: the fundamental
theory (or field theory, FT) and the effective field theory (EFT).
The latter takes only light states into account (the massless NS-NS
fields $G_{\mu\nu},B_{\mu\nu},\phi$) and computes correctly in the
low-energy range, {\em i.e.}~up to energies neighbouring the masses of
the heavy states (which occur at the scale of the string length).  At
tree-level, it computes the {\em on-shell} scattering amplitudes for
light states up to terms that vanish by the equations of motion, {\em
  i.e.}~it integrates out the heavy states.  The FT on the other hand
takes both light and heavy states into account; so the difference
between the FT and EFT result is precisely the contribution from heavy
states that the EFT misses out.  This contribution should be added to
the loop-expansion in the EFT and is called the {\em threshold
  correction} $\Delta_i$.  The subscript $i$ refers to the gauge group
for which we compute the gauge coupling.

At tree level, the gauge coupling constant is equal to the string
coupling.  Denoting the one-loop correction by $\Delta_i$, we have to
one-loop order:
\begin{equation}  \label{tree-level}
  \frac{1}{g_i^2}= \frac{k_i}{g^2_\str} + \Delta_i,
\end{equation}
The constant $k_i$ is the central element for the left-moving algebra
which generates the gauge group $G_i$.  Since $G_i$ is non-abelian, we
set $k_i=1$.  The threshold correction itself can be computed for an
N=2 heterotic compactification on $ K3 \times T^2 $ to be (see
\cite{G-04} for a sketch)
\begin{equation} \label{one-loop-IR-finite}
  \Delta_i = \int_{\CF}{d^2\tau\over \tau_2}  \left( {-i\over \pi
  |\eta|^4} \sum_{\rm even} (-1)^{a+b}
 ~\bar\p_{\bar\tau}\left({\bar\th[^a_b]\over \bar\eta}\right)~{\rm 
  Tr}_{\rm int}\left[Q_i^2-{k_i\over 4\pi\tau_2} \right][^a_b] -b_i \right).
\end{equation}
The subscript $i$ refers to one of the several factors of the gauge
group, while ``even'' spin structures means $(a,b)\neq (1,1) $
($a,b=0,1$).  The constant $b_i$ is called the {\em beta function
  coefficient} and represents the constant term of the $q$-expansion
of the integrand (see also appendix E of \cite{G-04}); subtracting it
renders the integral IR finite.  Note that we have suppressed for
convenience the factors of $q$ in the trace: the proper expression
should contain also $ C_{\rm int}[^a_b] := {\rm tr}~ (-1)^{b\bar J_0}
q^{L_0 -\frac{11}{12}} \bar q^{\bar L_0 -\frac{3}{8}} [^a_b] $ for the
$(c,\bar c)=(22,9)$ internal theory, where $\bar J_0$ coincides with
the right-moving fermion number $F_R$.  The notation $[^a_b]$ stands
for the spin structure, see also (\ref{C_int}).

\subsection{Threshold via New Susy Index and $K3$ Elliptic Genus}

An alternative way of writing the integrand (\ref{one-loop-IR-finite})
is as a ``new susy index'' (\ref{susy-index}) (see again \cite{G-04}
for a sketch) 
\begin{equation}  \label{one-loop-result-bis}
  {16\pi^2\over g_i^2}\bigg|_{\rm 1-loop}
  ={-1\over \eta^2} \int_{\CF}{d^2\tau\over \tau_2} \left( {\rm Tr}_{R, \rm
  int} \left( \bar F (-1)^{\bar F} q^{L_0 -\frac{11}{12}} \bar q^{\bar
  L_0 -\frac{3}{8}} \left[Q_i^2-{k_i\over 4\pi\tau_2} \right]\right) -
  b_i \right),
\end{equation}
with the usual group theory factor in square brackets.  Indeed, this
was the starting point in \cite{HM-95}.

The computation of this new susy index is quite different for the
left- and right-moving sectors.  In the right-moving sector, the
algebra factorises into the direct sum of a $\bar c=3$ N=2 SCA and a a
$\bar c=6$ N=4 SCA, which simplifies considerably the computation of
the new susy index: the result is a mere constant, $-2i$, thanks to
the equal but opposite contributions of vector- and hypermultiplets
towards $\bar J_0 (-1)^{\bar J_0}$ (see \cite{HM-95} or section 10.5
of \cite{G-04}).  In the left-moving sector, the new susy index enters
in the form of the above elliptic genus tr$_R ~(-1)^{J_0 +\bar J_0}
q^\Delta $, to which we now turn.

\subsubsection*{$K3$ Elliptic Genera}

The elliptic genus of a $(c,\bar c)=(6,6)$ heterotic sigma model on
$K3 \times T^2$ is, geometrically, a double sum $ \Phi (\tau,z) =
\sum_{n,r} c_{n,r} ~q^n y^r $ whose coefficients are the indices of
Dirac operators for certain vector bundles over $K3$.  The elliptic
genus also has a topological expression, given by a trace over the
left and right Ramond sectors with $(-1)^F$ insertion:
\begin{equation}\label{ell-genus}
  \begin{split}
    \Phi (\tau,z) &:= \tr_{_{\rm R,R}} (-1)^{F_L+F_R} q^{L_0-1/4} \bar
  q^{\bar L_0-\bar 1/4} y^{J_0} \\
  &= 24\left(\frac{\th_3(z)}{\th_3}
  \right)^2 +2 \frac{\th_2^4-\th_4^4}{\eta^4} \left(
   {\th_1(z)\over\eta} \right)^2  ,
  \end{split}
\end{equation}
where the second expression was proved in \cite{EOTY-89}, \cite{G-03},
and corresponds to the unique weak Jacobi form of weight 0 and index 1
(see appendix~\ref{sec:jacobi-forms}) with $\Phi(\tau,0)
=\chi(K3)=24$.  
At the special values of $z=\frac{1+\tau}{2}, \frac{\tau}{2},
\frac{1}{2}$ and $0$, we obtain specific topological invariants
\cite{EOTY-89}, using (\ref{t13}),(\ref{abstruse}) and dropping the
extra $q^{-1/4}$ and $-q^{-1/4}$ in the first two cases:
\begin{equation}  \label{top-invar}
  \begin{array}{llll}
{\rm Dirac\  index:}\qquad &\Phi^+_{\hat A} &:= \tr_{_{\rm NS,R}}
  (-1)^{F_R} q^{L_0-1/4} &= 2 \th_3^2
  (\th_2^4 -\th_4^4) / \eta^6 \\
&\Phi^-_{\hat A}&:= \tr_{_{\rm NS,R}} (-1)^{F_L+F_R} q^{L_0-1/4}  &= -2 \th_4^2
  (\th_2^4+\th_3^4)  / \eta^6 \\
{\rm Hirzebruch\ genus:}\qquad &\Phi_\sigma &:= \tr_{_{\rm R,R}} (-1)^{F_R}
  q^{L_0-1/4} &= 2 \th_2^2
  (\th_4^4+\th_3^4) / \eta^6  \\
{\rm Euler\  character:}\qquad &\Phi_\chi &:= \tr_{_{\rm R,R}}
  (-1)^{F_L+F_R} q^{L_0-1/4} &= 24
  \end{array}
\end{equation}
Whence a shift $z\to z+\frac{\tau}{2}$ generates spectral flow
R$\to$NS, while $z\to z+\frac{1}{2}$ is responsible for an additional
factor of $(-1)^{F_L}$.  The elliptic genus evaluated at specific
points thus yields the partition function for different spin
structures (boundary conditions in the left-moving sector); at $z=0$,
we obtain the Witten index -- or the bosonic partition function if we
have no spin structures. 

\subsubsection*{``New Susy Index''}

It remains us to compute explicitly the new susy index for the
left-moving sector, that is to evaluate the above elliptic genus or
to find the proper partition functions.

In our compactification on $T^2\times K3 $, we shall use bosonic
formulation for one of the $E_8$ gauge groups and fermionic
formulation (with sixteen left-moving fermions) for the other.  The
gauge bundle is a direct sum of a bundle on $T^2$ and one on $K3$,
with flat connection or a.s.d.~connection respectively.  We choose to
couple 12 fermions with the former connection and 4 with the latter,
so as to obtain the desired $(c,\bar c)=(6,6)$ heterotic sigma model
on $K3$.\footnote{In general, $(16-2n)$ fermions on $T^2$ plus $2n$ on
  $K3$ gives a $(c, \bar c) = (4+n, ~6)$ model.  Hence each fermion
  coupled to $K3$ increases the left-moving central charge by 1/2.  We
  have $n=2$.}  The partition functions for the former fermions on
$T^2$ are the familiar $\th_i/\eta, \ i=1,2,3,4$ for the NS/R sectors
with/without $(-1)^F$ insertion.  The bosonic realisation of the other
$E_8$ factor yields the familiar theta function $\Gamma_{8}/\eta^8$,
which we join with the $\Gamma_{2,2}$ lattice of $T^2$ to obtain
$\Gamma_{10,2}/\eta^{12}$.  Taking into account the $-2i$ from the
right-moving sector, and summing over all worldsheet boundary
conditions of the left-moving sector, we obtain for the whole ``new
susy index'' of the internal theory:
\begin{equation*}
  \begin{split}
   {\rm tr_{_R}} ~\bar J_0  &~e^{i \pi (J_0-\bar J_0)} q^{L_0-c/24} \bar
   q^{\bar L_0-\bar c/24} \\
  &= -2i \frac{\Gamma_{10,2}}{\eta^{12}} \left(
   \bigg(\frac{\th_3}{\eta}\bigg)^6 \Phi^+_{\hat A} +
   \bigg(\frac{\th_4}{\eta}\bigg)^6 \Phi^-_{\hat A} +
   \bigg(\frac{\th_2}{\eta}\bigg)^6 \Phi_\sigma +
   \bigg(\frac{\th_1}{\eta}\bigg)^6 \Phi_\chi \right) \\
 &=  -2i \frac{\Gamma_{10,2}}{\eta^{12}} 
  \frac{2}{\eta^{12}} \bigg( \th_3^8 (\th_2^4-\th_4^4)
   -\th_4^8 (\th_2^4+\th_3^4) +\th_2^8 (\th_3^4+\th_4^4)\bigg) ,
  \end{split}
\end{equation*}
hence   
\begin{equation}  \label{new-susy-index-concrete}
 {\rm tr_{_R}} ~\bar J_0  ~e^{i \pi (J_0-\bar J_0)} q^{L_0-c/24} \bar
   q^{\bar L_0-\bar c/24} ~~= 8i\ \Gamma_{10,2}\ \frac{E_6}{\Delta},  
\end{equation}
where we noted that the last bracket is but $\sum_{i\neq j}\th_i^8
\th_j^4$ with a minus sign if $2i+j>8 $, which is just $-2 E_6$ by
(\ref{E4}).

\subsection{Threshold via Heterotic Orbifold Partition Function} 
\label{sec:het-orb-part-fct}

We will now derive the same result for the ``new susy index'', but this
time starting from the expression (\ref{one-loop-IR-finite}) for
the partition function
\begin{equation}  \label{het-partition-fct}
  Z^{\rm het}_{D=4}={1\over \tau_2\eta^2\bar \eta^2}\sum_{a,b=0}^1
~(-1)^{a+b+ab}~{\bar\th[^a_b]\over \bar\eta}~C_{\rm int}[^a_b]
\end{equation}
with the appropriate internal contribution $ C_{\rm int}[^a_b] = {\rm
  tr}_{\rm int}~ (-1)^{b\bar J_0} q^\Delta \bar q^{\bar \Delta} [^a_b]
$, but still w ithout the gauge theory factor with the Casimir
operator.  It i s encouraging to see the alternative result
(\ref{integrand-result}) agree with (\ref{new-susy-index-concrete}).

The internal contribution consists of the partition functions for the
$T^2$ bosons ($\Gamma_{2,2}(T,U)/\eta^2\bar\eta^2$) and fermions ($\half
\sum_{a,b} (-1)^{a+b+ab} \bar\th[^a_b]/\bar\eta $), for the $K3$
bosons and fermions, as well as for the gauge bundle and its two $E_8$
factors which we take in the bosonic realisation ($\Gamma_8/\eta^8 =
E_4/\eta^8$) and fermionic realisation
($\half\sum_{\gamma,\delta}\th^8 [^\gamma _\delta]/\eta^8 $)
respectively.  Since the partition function (or the elliptic genus) is
a topological object, it does not depend on the hypermultiplet
moduli, so we will choose a limit for these moduli where the $K3$
surface is described by a $\Z_2$ orbifold breaking the gauge group to
$ E_8 \times E_7 \times SU(2)$.  This has a well-known partition
function for the bosonic (4,4) blocks:
\begin{equation}\label{bosonic-blocks}
 Z_{(4,4)}[^0_0]=Z(R)= {\Gamma_{4,4}(G,B)\over
\eta^4\bar\eta^4},\qquad Z_{(4,4)}[^h_g]=2^4{\eta^2\bar
\eta^2\over \th^2[^{1-h}_{1-g}]\bar\th^2[^{1-h}_{1-g}]}
\qquad (h,g)\neq(0,0),
\end{equation}
where the lattice function depends on the metric $G_{ij}$ and the
B-field $B_{ij}$:
\begin{equation}\label{lattice-function}
  \begin{split}
    {\Gamma_{4,4}(G,B)\over \eta^4\bar\eta^4}
 &:= \sum_{m,n\in\Z^4} {q^{p_L^2/2} \bar q^{p_R^2/2}\over\eta^4\bar\eta^4}, 
  \qquad  p_{L,R}^i:= {G^{ij}\over\sqrt{2}}(m_j+(B_{jk}\pm G_{jk})n_k) \\
 &= {\sqrt{{\rm det}~G}\over(\sqrt{\tau_2} \eta\bar\eta)^4} \sum_{m,n\in\Z^4} 
   \exp \left( -{\pi\over\tau_2}
 (G_{ij}+B_{ij})(m_i+n_i\tau)(m_j+n_j\bar\tau) \right) 
  \end{split}
\end{equation}
with  $p_{L,R}^2$ the inner product {\em w.r.t.}~the metric, {\em i.e.}~$p_{L,R}^i
G_{ij} p_{L,R}^j$.  (Note that our metric $G_{ij}$ has absorbed a
factor of $R_i$ (orbifold radii) compared to other conventions.)

Similarly, the compact $K3$ fermions ($\bar\th^2/\bar\eta^2$)
are twisted in the (4,4) blocks: $\bar\th[^{a+h}_{b+g}]
\bar \th[^{a-h}_{b-g}] / \bar\eta^2$.  Additionally, the orbifold
projection on the fermionic $E_8$ factor will correspond to a sign
change for the doublet of the $SU(2)$ subgroup of $E_8 \supset  E_7
\times SU(2)$.  Note that the adjoint decomposes as 
$$
{\bf 248 \to (133,1) + (1,3) + (56,2)} \in E_7 \times SU(2),
$$ 
and that the projection acts on the $SU(2)$ representations as
${\bf 3 \to 3, \quad 2\to -2}$.  This entails that two of our eight
complex fermions are twisted by the projection and this part of the
partition function reads then
\begin{equation}\label{E_7}
  \half\sum_{\gamma,\delta=0}^1~{\th[^{\gamma+h}_{\delta+g}]\th[^{\gamma-h}_{\delta-g}]\th^6[^{\gamma}_{\delta}]\over\eta^8}.
\end{equation}
The other $E_8$ factor (the bosonic realisation with $\Gamma_8$) is
not affected by the projection. The full partition function for the
internal theory is the product of all the above-mentioned parts:
\begin{equation} \label{C_int}
  C_{\rm int} = 
  {\Gamma_{2,2}\over \eta^2\bar\eta^2} {\Gamma_8\over \eta^8} 
  \ \half \sum_{a,b}(-1)^{a+b+ab}{\bar\th [^a_b]\over \bar\eta} 
  \ \half \sum_{g,h=0}^1 Z_{(4,4)}[^h_g] {\bar\th[^{a+h}_{b+g}]
  \bar\th[^{a-h}_{b-g}]\over \bar\eta^2}
  \ \half \sum_{\gamma,\delta=0}^1 ~{\th[^{\gamma+h}_{\delta+g}]
  \th[^{\gamma-h}_{\delta-g}] \th^6[^{\gamma}_{\delta}] \over\eta^8}  
\end{equation}
In (\ref{het-partition-fct}) or implicitly in
(\ref{one-loop-IR-finite}), $C_{\rm int}[^a_b]$ is just the above with
$ C_{\rm int} =: \sum_{a,b} ~(-1)^{a+b+ab} ~C_{\rm int}[^a_b] $.

Note that the term in (\ref{one-loop-IR-finite}) with the $
\bar\tau$-derivative includes the non-compact fermions.  We will combine
these with the $T^2$ and $K3 $ fermions and sum over their spin
structures:
\begin{equation*}
  \begin{split}
    \frac{-i}{\pi} \half \sum_{\rm even} (-1)^{a+b}
  & ~\bar\p_{\bar\tau}\left({\bar\th[^a_b]\over \bar\eta}\right)
   ~{\bar\th[^a_b]\bar\th[^{a+h}_{b+g}] \bar\th[^{a-h}_{b-g}]\over\bar\eta^3}\\
   &=\frac{1}{24 \bar\eta^4} \bigg[
   (\bar\th_2^4-\bar\th_4^4)\bar\th_3^2\bar\th[^h_g]\bar\th[^{-h}_{-g}] 
  - (\bar\th_3^4+\bar\th_4^4)\bar\th_2^2\bar\th[^{1+h}_g]\bar\th[^{1-h}_{-g}]
  + (\bar\th_2^4+\bar\th_3^4)\bar\th_4^2\bar\th[^h_{1+g}]\bar\th[^{-h}_{1-g}]
  \bigg]  \\
  &= -\half \bar\eta^2 \bar\th[^{1-h}_{1-g}]^2 (-1)^{1+h}
  \end{split}
\end{equation*}
where we have used (\ref{t16-tris}) and (\ref{t13}); and the use of
(\ref{abstruse}) will convince the reader that whatever the
combination of $g$ and $h$, the square brackets yield $ -12~\bar\eta^6
\bar\th[^{1-h}_{1-g}]^2 (-1)^{1+h}$. In particular, this vanishes for
$(g,h)=(0,0)$.  

If we multiply it with the $K3$ bosons and the uncompactified bosons
($1/\eta^2\bar\eta^2$ for two transverse bosons in the LCG), we
obtain, noting that $ \th[^{1-h}_{1-g}] (-1)^{1+h}= \th[^{1+h}_{1+g}]$: 
\begin{equation} \label{K3-bosons}
-8 ~\frac{\bar\eta^2}{\th[^{1+h}_{1+g}]\th[^{1-h}_{1-g}]}, \qquad
 (h,g)\neq(0,0).
\end{equation}
Taking further into account the remaining $E_8$ fermions with orbifold
projection, and summing explicitly over the $g,h$ blocks of the orbifold:
\begin{equation} \label{sum-g-h}
  \begin{split}
    -8 \frac{\bar\eta^2}{\eta^8} \frac{1}{4} \sum_{(h,g)\not=(0,0)}
    ~\sum_{a,b=0}^1
& {\th[^a_b]^6 \th[^{a+h}_{b+g}]\th[^{a-h}_{b-g}]\over
   \th[^{1+h}_{1+g}]\th[^{1-h}_{1-g}]} 
= -8 \frac{\bar\eta^2}{\eta^8} \frac{1}{4}\left(
  \frac{\th_2^6\th_4^2 -\th_4^6\th_2^2}{\th_3^2}  
   + \frac{\th_3^6\th_2^2+\th_2^6\th_3^2}{\th_4^2}
   - \frac{\th_3^6\th_4^2 +\th_4^6\th_3^2}{\th_2^2} \right) \\
&=  -8 \frac{\bar\eta^2}{\eta^8} \frac{1}{4} \frac{1}{4\eta^6}\bigg(
  \th_2^4\th_4^4 (\th_2^4-\th_4^4) +\th_2^4\th_3^4 (\th_2^4+\th_3^4) 
   - \th_3^4\th_4^4 (\th_3^4+\th_4^4) \bigg) \\
&= \frac{\bar\eta^2}{\eta^{14}} ~E_6
  \end{split}
\end{equation}
where the last bracket yields $-2 ~E_6$ according to (\ref{E4}). 

Finally, combining the non-projected $E_8$ bosons with the $T^2$
bosons gives us the lattice function $\Gamma_{10,2}/\eta^{10}\bar
\eta^2 $, and we obtain for the overall expression of the integrand of
(\ref{one-loop-IR-finite}) without the gauge theory factor: 
\begin{equation}\label{integrand-result}
  {-1\over 4\pi^2} {1\over \eta^2\bar\eta^2} \sum_{\rm even} (-1)^{a+b} 4\pi
i~\bar\p_{\bar\tau}\left({\bar\th[^a_b]\over \bar\eta}\right) ~C_{\rm int}[^a_b]= \frac{\Gamma_{10,2} ~E_6}{\eta^{24}} =\Gamma_{2,2}\frac{E_4~E_6}{\eta^{24}}.
\end{equation}
This is just the ``new susy index'' (\ref{new-susy-index-concrete}),
obtained in a different approach (SCFT for a sigma-model) but with the
gauge bundle still split into one factor with bosonic realisation and
the other with fermionic one.  Thus our explicit example of $T^2
\times K3$ with its two independent calculations is a verification
that the integrands of (\ref{one-loop-IR-finite}) and
(\ref{one-loop-result-bis}) are indeed the same!

To give a feeling of the information content of these partition
functions about the orbifold theory at hand, we show how the massless
spectrum can be arrived at.  We use the partition functions in
(\ref{C_int}) and claim that the twisted sector contains the massless
states 8 ({\bf 56,1,1}) and 32 ({\bf 1,2,1}) in the $E_7 \times SU(2)
\times E_8$.  The remaining ({\bf 56,2,1}) +4 ({\bf 1,1,1}) are in the
untwisted sector.  Now the twisted sector in (\ref{C_int}) corresponds
to $h=1$ and the bosonic part of it is given by $a=0$ (for which we
have $\th_3,\th_4$, {\em i.e.}~half-integer powers of $q$, {\em
  i.e.}~the NS sector).  The right-moving fermions (including two
non-compact transverse fermions $\th[^a_b]/\eta$) give us
\begin{equation} \label{right-moving-fermions}
  {1\over 2\bar\eta^4}
  (\bar\th^2_3\bar\th^2_2+\bar\th^2_4\bar\th^2_2)
  \qquad (g=0 ~{\rm and } ~g=1~ {\rm resp.}).
\end{equation}
Using this expression, we look at the lowest powers (massless states)
in the full partition function 
\begin{equation*}
  \begin{split}
    \half \sum_{a=0,b} ~\half \sum_{g,h=1} \dots Z_{4,4}[^h_g]
  ~\half \sum_{\gamma,\delta} \dots
  &= {1\over 4} \bar\th^2_2 |\th_2|^4 \left[ \bar\th_3^2 |\th_3|^4
  (\th_2^2\th_3^6 +\th_3^2\th_2^6)  +\bar\th_4^2 |\th_4|^4 (\th_3^2\th_2^6
  -\th_2^2\th_4^6) \right] \\
  &= {1\over 4} ~2^4 \bar q^{1/2}(1+\cdots) ~4 q^{1/4}(1+\cdots) 
  \left[ 4 q^{1/4} (16 \bar q^{1/2}+\cdots) \right] \\
    &= 2^{10} q \bar q^{1/2} +\cdots
  \end{split}
\end{equation*}
So we have $2^{10}$ twisted massless scalars, which is the bosonic
content of $2^{9}$ twisted massless hypermultiplets, which is just
$8\cdot {\bf 56} +32\cdot {\bf 2}$ as expected from the spectrum.

\subsection{Examples of Threshold Corrections}
\label{sec:examples-threshold-corr} 

In section \ref{sec:het-orb-part-fct}, we devised a convenient way for
calculating the integrand of the one-loop threshold corrections, by
combining the ingredients of the partition function of the heterotic
orbifold compactification.  We will now proceed to incorporate the
gauge theory factor, that is the trace of the Casimir operator or the
square brackets in (\ref{one-loop-IR-finite}).

Let us go back to our N=2 $\Z_2$ orbifold $T^2\times K3$ of
section~\ref{sec:het-orb-part-fct} with gauge group $E_8 \times E_7
\times SU(2)$.  We shall let $\left[Q_i^2-{k_i\over 4\pi\tau_2}
\right]$ act on the three factors of the gauge group separately, by
replacing $Q_i^2$ with $(\frac{\p_{v_1}}{2\pi i})^2 |_{v_j=0} $ acting
on the character $\chi_R (v_j)$ of the corresponding factor.  That is,
we first write the character with a $v$-dependence and then
differentiate twice {\em w.r.t.}~only one of the $v_j$, say $v_1$.
For instance, the $E_8$ character is $\chi_{E_8}=\Gamma_8/\eta^8 =
E_4/\eta^8 = (\th_2^8+\th_3^8+\th_4^8)/2\eta^8 $, and with
$v$-dependence it becomes
\begin{equation} \label{v-dep}
\chi_{E_8} (v_j) = \half \sum_{a,b=0}^1 {\prod_{j=1}^8
  \th[^a_b](v_j)\over \eta^8} . 
\end{equation}
When acting with $\left[(\frac{\p_{v_1}}{2\pi i})^2 |_{v_j=0}
-{k_i\over 4\pi\tau_2} \right]$, the operator on the left can be
replaced by $ {1\over i\pi}\p_{\tau}$, see (\ref{t16}), with
an extra factor of $1/8$ since the other theta functions are now also
affected by the $\p_\tau$.  Taking out this factor of  $1/8$, we are
left with the covariant derivative $D_4$ of (\ref{E10-bis}), and the
whole expression is, by (\ref{cov-deriv}):
\begin{equation*}
 \left[(\frac{\p_{v_1}}{2\pi i})^2 |_{v_j=0}
-{k_i\over 4\pi\tau_2} \right] \chi_{E_8} (v_j) = {\hat E_2~E_4 - E_6
\over 12 ~\eta^8}. 
\end{equation*}
Since all other factors of the gauge group or of the internal
partition function remain the same, we simply ought to replace the
$\Gamma_8/\eta^8$ factor of (\ref{integrand-result}) by the above
result, and the complete threshold (\ref{one-loop-IR-finite}) is:
\begin{equation} \label{threshold-E_8}
  \Delta_{\rm E_8}=\int_{\CF}{d^2\tau\over \tau_2}\left[-{1\over
12}\Gamma_{2,2}{\hat{ E}_2 E_4 E_6- E_6^2\over \eta^{24}}+60\right],
\end{equation}
where we bear in mind that, as before, the lattice function
$\Gamma_{2,2}$ depends on the torus moduli: $T,U \in \CH$.  We note
that the (nearly) modular form in the fraction has a Fourier expansion
starting with $720+\dots$, that $\Gamma_{2,2}$ starts with $1+\dots$,
so that dividing by $-12$, we do indeed obtain $-60$ as the constant
term.  This agrees with the beta function coefficient $b_{E_8}$ of the
corresponding orbifold (see appendix E.2 of \cite{G-04}).

Similarly, for the threshold corresponding to the $E_7$ factor of the
gauge group, we exchange the roles of the bosonic and fermionic $E_8$
partition functions.  That is we keep $\Gamma_8/\eta^8$ unchanged
and let $\left[(\frac{\p_{v_1}}{2\pi i})^2 |_{v_j=0} -{k_i\over
  4\pi\tau_2} \right]$ act on the $E_7$ character (\ref{E_7}) with
$v_1$-dependence:
\begin{equation}\label{fermionic}
 \chi_{E_7} (v_1) =  {1\over 2}\sum_{a,b}{\th[^a_b](v_1)\th^5[^a_b]
\th[^{a+h}_{b+g}]\th[^{a-h}_{b-g}]\over \eta^8}
\end{equation}
We remember from (\ref{sum-g-h}) that these were combined with
$\bar\eta^2/\th\th$ of (\ref{K3-bosons}) and summed over $g,h$ to
yield $E_6$.  This is basically a product of 12 theta functions, of
which only the first will have the $v_1$ dependence, so that we can
replace the $\p_{v_1}^2$ by ${1\over 12} \p_\tau$ to have $D_6~E_6$,
again according to (\ref{t16}).  The covariant derivative yields
$E_4^2-\hat E_2~E_6$, and combining this with the remaining toroidal
bosons $\Gamma_{2,2}$ and $\Gamma_8/\eta^8$, we obtain the desired
threshold corrections:
\begin{equation}\label{threshold-E_7}
  \Delta_{\rm E_7} =\int_{\CF}{d^2\tau\over \tau_2}\left[-{1\over
12}\Gamma_{2,2}{\hat{E}_2 E_4 E_6- E_4^3\over \eta^{24}}-84\right] ,
\end{equation}
where the fraction has a Fourier expansion starting with
$-1008+\dots$, which yields 84 when divided by 12.  This again
agrees with the beta function coefficient $b_{E_7}$ of the
corresponding orbifold.

Note that (\ref{fermionic}) contains also the character for the
$SU(2)$ subgroup; so that if we were to compute the trace of the
$SU(2)$ Casimir, we would obtain the same result for the threshold and
consequently for $b_{SU(2)}$: 84, also agreeing with the table of
beta-function coefficients in appendix E.2 of \cite{G-04}.
 
It is interesting to note that the difference of our two thresholds
can be easily computed: recalling that $\eta^{24}=
\Delta=(E_4^3-E_6^2)/1728$, we evaluate the integral using the trick
displayed in \cite{DKL-91} (``lattice unfolding technique''):
\begin{equation*}
  \Delta_{\rm E_8}- \Delta_{\rm E_7} = -144 \int_{\CF}{d^2\tau\over
  \tau_2} (\Gamma_{2,2}-1) 
  = 144 \log \left(4\pi^2 T_2 U_2|\eta(T)\eta(U)|^4 \right) ,
\end{equation*}
which scales as $\sim T_2$, {\em i.e.}~as the volume of the torus, in the
decompactification limit. 

The above case of N=2 orbifold is closely related to the N=1 orbifold
where we introduce a second $\Z_2$ twist to obtain a $\Z_2 \times
\Z_2$ orbifold with gauge group $E_8 \times E_6 \times U(1)^2$.
However, the construction is independent of the untwisted moduli
$(T_i,U_i)$ of the three twisted 2-planes, and the threshold
corrections are not affected by this N=1 sector.  So the threshold
corrections carry over from the N=2 sectors (only one 2-plane
twisted): $ \Delta_{\rm E_8} $ is as in (\ref{threshold-E_8}) and $
\Delta_{\rm E_6}$ as in (\ref{threshold-E_7}).  Consequently, the
constant term of the whole integrand is 3/2 of what it used to be (3
for the three 2-planes and 1/2 due to the extra $\Z_2$ twist).

The reader will wonder what the above explicit expressions for
one-loop threshold corrections to gauge couplings have to do with the
Gromov-Witten invariants we started with in the first chapter.  The
answer was in fact already given in section \ref{sec:result-prepot}:
these $\Delta_i$ occur in an ODE for the prepotential $F_0$, {\em
  i.e.}~the genus-0 GW potential.  The resolution of the integrals
over the fundamental domain yield polylog expressions that can be
reorganised as in (\ref{hm}).  This convenient sum allowed
\cite{HM-95} to extract the coefficients $c(- {r^2 \over 2})$, with
meaning in enumerative geometry as GW invariants generally have.

\appendix
\newpage
\section{Appendix: Jacobi forms}\label{sec:jacobi-forms}

\paragraph{Definition}

According to \cite{EZ-85}, a (holomorphic) {\bf Jacobi form of weight $k$ and
index $m$} (non-negative integers) is a holomorphic function
$\phi(\tau,z): \CH \times \C \mapsto \C $ satisfying the following
three conditions:
\begin{equation*}
  \begin{split}
    \phi \Big(\frac{a\tau+b}{c\tau+d},\frac{z}{c\tau+d} \Big) &=
    (c\tau+d)^k e^{2\pi i\frac{mcz^2}{c\tau+d}} \phi(\tau,z),
    \hspace{3cm} {{a \quad b} \choose {c \quad d}} \in SL(2,\Z), \\ 
    \phi(\tau,z+\lambda\tau+\mu) &= e^{-2\pi i (\lambda^2\tau
    + 2\lambda z)}  \phi(\tau,z), \hspace{3cm}  \lambda,\mu
    \in\Z,\\
     \phi(\tau,z) &= \sum_{n\geq 0} \sum_{\substack{r \in
    \Z \\ r^2\leq 4nm}} c_{n,r} \ q^n y^r, \hspace{3cm} q=e^{2\pi
    i\tau}, y=e^{2\pi iz}.
  \end{split}
\end{equation*}
The first condition is reminiscent of transformation properties of
theta functions; on the divisor $z=0$, $\phi(\tau,0)$ is a modular
form of weight $k$.  A remarkable property is that this extends to
rational portions $z$ of the fundamental lattice $\{1,\tau \}$ :
$e^{2\pi im\lambda^2\tau} \phi(\tau,\lambda\tau+\mu)$ (with
$\lambda,\mu \in\Q$) is a modular form of weight $k$ --if not
identically zero--, though in general only for a subgroup of $
SL(2,\Z)$.  The accompanying behaviour of $z\mapsto \frac{z}{c\tau+d}$
follows from the postulated behaviour under the two $SL(2,\Z)$
generators $S$ and $T$: $\phi(-1/\tau,z/\tau)$ and $ \phi(\tau
+1,z) $. 

\paragraph{Periodicity}

The second condition similarly reproduces periodicity properties of
theta functions.  It entails the periodicity of the coefficients:
$$
c_{n,r}=c_{n',r'} \hspace{3cm} \textrm{ if } r'\equiv r ~(\textrm{mod}~
2m) \quad \textrm { and } n'=n+\frac{{r'}
^2 - r^2}{4m},
$$ 
or equivalently
\begin{equation}  \label{periodicity}
c_{n,r}=c_{n+m-r,~r-2m}=c_{n+m+r,~r+2m},
\end{equation}
a hallmark of Jacobi forms.

\paragraph{The condition ${\bf r^2\leq 4nm}$:}

The third condition is an interesting one; first it requires the
convergence of the Fourier series, secondly it requires $r^2\leq 4nm$.
The latter requirement can be explained from three different
perspectives.  

Firstly, it is a consequence of the above-mentioned fact that $e^{2\pi
  m i\lambda^2\tau} \phi(\tau,\lambda\tau)$ is a modular form for all
$\lambda\in\Q$: holomorphicity at $\tau =\infty$ requires
$n+r\lambda+m\lambda^2 \geq 0 \quad \forall \lambda$, {\em i.e.}~$r^2\leq
4nm$ after replacing $\lambda$ by the extremum.

Secondly, $r^2\leq 4nm$ is the condition for the matrix
$T={n\quad r/2 \choose r/2\quad m}$ to be positive semi-definite,
$m,n,r\in\Z$.  If we sum over such semi-integral $2\times2$ matrices
in the Fourier-expansion
$$
F(Z):=\sum_{T\geq 0} c_T \ e^{2\pi i\ \textrm{tr }TZ},
$$ 
we obtain a so-called {\it Siegel modular form} of degree 2, that
is a holomorphic function of the generalised upper-half plane $\CH_2$
of all symmetric $2\times2$ matrices with positive definite
imaginary part, {\em i.e.}~${\tau \quad z \choose z \quad \tau'}$
with $\tau, \tau'\in\CH, z\in\C$, covariant under the action
of the Siegel modular group $Sp(4, \Z)$:
$$
F\Big(\frac{AZ+B}{CZ+D}\Big)= \textrm{det} (CA+D)^k \ F(Z), \qquad 
\left( \begin{array}{cc}
  A &B\\
  C&D
\end{array}  \right)  \in Sp(4,\Z)
$$
for some weight $k$.
Note that tr $TZ =n\tau+rz+m\tau'$ and that we may as well rewrite
the Fourier expansion as
$$
F(\tau,z,\tau')= \sum_{m\geq 0} \Big
( \sum_{\substack{n,r\in\Z \\ r^2\leq 4nm}} c_T \ q^n y^r \Big)
e^{2\pi im\tau'} ,
$$
in which the bracketed piece is a Jacobi form of weight $k$ and index
$m$.  In this larger context of Siegel modular forms, we see the
origin of the condition $r^2\leq 4nm$.

A third way to understand the condition is through an example of a
Jacobi form: a theta function, defined by summing over the roots $x$
of an even, self-dual lattice $\Gamma$ of rank $N$:
$$
\phi(\tau,z):= \sum_{x\in\Gamma} q^{x^2/2} y_0^x, \qquad
y_0^x=e^{2\pi i (x_0\cdot x)z}, x_0\in\Gamma.
$$
This is a form of weight $k=N/2$ and index $m=x_0^2/2$, and the Schwarz
inequality for the bilinear form associated to $\Gamma$ reads
$(x\cdot x_0)^2 \leq x^2 x_0^2$, that is $r^2\leq 4nm$.  For more
general lattices (not necessarily self-dual), the form will be
covariant under only a congruence subgroup of $SL(2,\Z)$.

\paragraph{Zeros}

For fixed $\tau$, a simple contour integration shows that the the
number of zeros minus the number of poles in a fundamental domain for
$\C / (\Z\tau + \Z)$ is given by $2m$.  Since our Jacobi forms are
assumed holomorphic, this is just the number of zeros.  If $m$
vanishes, the second condition requires $\phi$ to be doubly-periodic
--hence constant-- in $z$, that is, $\phi$ is a mere modular form.  Had
we allowed $m$ to be negative, we would have {\it meromorphic} Jacobi
forms -- examples of which are inverses of Jacobi forms.

\paragraph{Decomposition}

The periodicity of the coefficients entails that we can write a Jacobi
form as a finite linear combination of theta functions with modular
forms as coefficients.  Indeed, (\ref{periodicity}) tells us that the
$c_{n,r}$ only depend on $4mn-r^2$ and on $r$ (mod $2m$).  Thus, for a
fixed residue class $\mu\in \Z/2m\Z$, all $c_{n,r}$ with $ r
\equiv\mu$ (mod $2m$) only depend on $N:=4mn-r^2$; for these we define
$c_\mu(N):=c_{4nm-r^2,r}$.  We then have $2m$ functions
\begin{equation*}
  \begin{split}
h_\mu(\tau):= & \sum_{N\geq 0} c_\mu(N)\ q^{N/4m}, \qquad \mu\in
\Z/2m\Z\\
    = & q^{-\mu^2/4m} \sum_{n\geq 0} c_{n,\mu}\ q^n \ ,
  \end{split}
\end{equation*}
wherein it is understood that $c_\mu(N)=0$ if $N\not\equiv -\mu^2$ (mod
$4m$).  These are vector-valued modular forms of weight $k-1/2$ (with
$2m$ components), which, under modular transformations, transform into
linear combinations of themselves.  They allow us to rewrite the
Jacobi form $\phi$ in a basis of theta functions:
\begin{equation}\label{decomposition}
  \begin{split}
    \phi(\tau,z) = \sum_{\mu(\textrm{mod }2m)} h_\mu(\tau) \
    \theta_{m,\mu}(\tau,z) \\
    \theta_{m,\mu}(\tau,z):= \sum_{\substack{r\in\Z \\
    r\equiv\mu~(\textrm{mod }2m)}} q^{r^2/4m} y^r
  \end{split}
\end{equation}
The theta functions similarly transform under the Jacobi group into
linear combinations of themselves, with weight $1/2$ and index $m$ (so
they are not Jacobi forms strictly speaking).  This decomposition of
Jacobi form holds also for forms of half-integer weight (shall not
deal with them), as well as for weak Jacobi forms (see below).  In the
latter case $h_\mu$ may contain negative (fractional) powers of $q$,
and $N\geq 0$ ought to read $N\geq -m^2$; the $c_\mu(N)$ vanish
anyway when $N<-r^2$ (where $|r|\leq m, ~r\equiv \mu$ (mod $2m$)).

\paragraph{Weak Jacobi forms}

If the requirement $r^2\leq 4nm$ is dropped, one can still prove that the
coefficients $c_{n,r}$ vanish unless $r^2\leq 4nm+m^2$, in which case
we have a so-called {\em weak} Jacobi form (still holomorphic).  The
periodicity of the coefficients still holds, as does the property of
decomposition into a linear combination of theta function.

In the weak case, the weight $k$ can be negative but no less than
$-2m$ ($m$ still assumed positive).  If this holds, $\phi(\tau,0)$
must vanish identically, as there are no holomorphic modular forms of
negative weight (safe perhaps for proper subgroups of $SL(2,\Z)$).
The same is true for Jacobi forms of odd weight, whether weak or not.
In fact, the space of weak Jacobi forms of weight $k$ and index $m$ is
isomorphic to a direct sum of vector spaces of modular forms of
different weights: $M_k \oplus \ldots\oplus M_{k+2m}$ for $k$ even,
while for $k$ odd the subscripts run from $k+1$ to $k+2m-3$.  (Recall
that $M_k=\emptyset$ for $k$ odd.)  Better still, for $k$ even the
correspondence permits us to write a weak Jacobi form as
$$
\phi_{k,m}=\sum_{i=0}^m f_i ~A_{-2,1}^i ~B_{0,1}^{m-i}
$$
for two generators $A_{-2,1} , B_{0,1}$ of the ring of weak Jacobi
forms of even weight (which is thus a polynomial algebra over $M_*$),
and $f_i \in M_{k+2i}$.  Note that since $M_0$ contains the constant
functions while $M_2$ is empty, we see that $\phi_{0,1}$ is unique (up
to a constant) and equals $B_{0,1}$.  Furthermore,
$$
A_{-2,1} = -\frac{\vartheta_1^2(\tau,z)}{\eta^6(\tau)} \hspace{2cm}
\textrm{and} \hspace{2cm} B_{0,1} = -\frac{3}{\pi^2} \wp A_{-2,1}
$$ 
where $\vartheta_1$ is the first Jacobi theta function, with a simple
zero at $z=0$ ($\vartheta_1^2$ {\em transforms} as a Jacobi form of weight
1, index 1, but with multiplier $-i$), and $\wp$ is the Weierstrass
function, itself a meromorphic Jacobi form of weight $-2$ and index 0
(with a double pole at $z=0$).  $A$ and $B$ are both constant
functions at $z=0$, equal to 0 and 12 respectively.

See appendix A of \cite{G-04} for a fuller treatment.

\section{Appendix: Theta Functions and Modular Forms}\label{sec:app-theta}

This is drawn in part from the exhaustive Appendix A and F of  \cite{K-97}.

\centerline{\bf Theta functions}

\begin{equation}  \label{t6}
  \begin{array}{ll}
\th_1(v|\tau)= -i \sum_{n\in\Z} (-1)^n q^{(n+\half)^2/2} y^{n+\half} 
   &= 2q^{1\over 8}\sin[\pi v]\prod_{n=1}^{\infty}
   (1-q^n)(1-q^n y)(1-q^n y^{-1})\\
\th_2(v|\tau)=\sum q^{(n+\half)^2/2} y^{n+\half}
&=2q^{1\over 8}\cos[\pi v]\prod (1-q^n)(1+q^n y)(1+q^n y^{-1})\\
\th_3(v|\tau)= \sum q^{n^2/2} y^n
&=\prod (1-q^n)(1+q^{n-1/2} y)(1+q^{n-1/2} y^{-1})\\
\th_4(v|\tau)= \sum (-1)^n q^{n^2/2} y^n 
&=\prod (1-q^n)(1-q^{n-1/2} y)(1-q^{n-1/2} y^{-1})
  \end{array}
\end{equation}

\vskip .5cm
\centerline{\bf $v$-periodicity formulae}

\begin{equation*}
  \begin{array}{rl}
  \th[^a_b](v+\half) &=  \th[^a_{b-1}](v) \\
\th[^a_b](v+{\tau\over 2}) &=  i^b q^{-1/8} y^{-1/2}~\th[^{a-1}_b](v) \\
\th[^a_b](v+{1+\tau\over 2}) &= -i^{b+1} q^{-1/8} y^{-1/2}~\th[^{a-1}_{b-1}](v)\\
  \th[^a_b](v+1) &= (-1)^a ~\th[^a_b](v) \\
\th[^a_b](v+\tau) &=  (-1)^b q^{-1/2} y^{-1}~\th[^a_b](v)
  \end{array}
\end{equation*}

\vskip .5cm
\centerline{\bf Useful identities}

\begin{equation} \label{theta-eta}
  \th_2 = 2~{\eta(2\tau)\over\eta} \qquad  \qquad 
  \th_3 = {\eta^5\over\eta(2\tau)^2~\eta({\textstyle {\tau\over 2}})^2}  \qquad    \qquad 
  \th_4 = {\eta({\textstyle {\tau\over 2}})^2\over\eta}
\end{equation}
\begin{equation} \label{t13}
  \begin{array}{rl}
\th_2 \th_3 \th_4 &= 2~\eta^3 \\  
\th_3(z|\tau)~\th_3(z'|\tau) + \th_2(z|\tau)~\th_2(z'|\tau)
&= \th_3({z+z'\over 2}|{\tau\over 2})~\th_3({z-z'\over 2}|{\tau\over 2}) \\
\th_3(z|\tau)~\th_3(z'|\tau) - \th_2(z|\tau)~\th_2(z'|\tau)
&= \th_4({z+z'\over 2}|{\tau\over 2})~\th_4({z-z'\over 2}|{\tau\over 2}) \\
\th_2(v|\tau)^4-\th_1(v|\tau)^4 &=\th_3(v|\tau)^4-\th_4(v|\tau)^4
  \end{array}
\end{equation}
For $v=0$, the latter is but Jacobi's {\em abstruse identity}: 
\begin{equation} \label{abstruse}
  \th_3^4 =\th_2^4 +\th_4^4.  
\end{equation}

\vskip .5cm
\centerline{\bf Heat equation}

\be
\left[{1\over (2\pi i)^2}{\partial^2\over \partial v^2}-{1\over
i\pi}{
\partial\over \partial\tau}\right]\th[^a_b](v|\tau)=0,
\label{t16}\ee
as well as
\be \label{t16-tris}
\partial_{\tau}\log \frac{\th_2}{\eta}= \frac{i\pi}{12}(\th_3^4+\th_4^4)
\,,\ee
and more generally for $(a,b)\neq (1,1)$:
\be
\partial_{\tau}\log \frac{\th[^a_b]}{\eta}=
\frac{i\pi}{12}\left( \th^4[^{a+1}_b]-\th^4[^a_{b+1}] +(-1)^b
\th^4[^{a+1}_{b+1}] \right) 
\,.\ee

\vskip .5 cm
\centerline{\bf Modular Forms}

We list the first few Eisenstein series:
\be
E_2={12\over i\pi}\partial_{\tau}\log\eta = 1-24\sum_{n=1}^{\infty}
{nq^n\over 1-q^n} = 1-24 \sum_{n=1}^{\infty} {q^n\over (1-q^n)^2}
\,,\label{E2}\ee
\be
E_4={1\over 2}\left(\vartheta_2^8+\vartheta_3^8+\vartheta_4^8\right)
=1+240\sum_{n=1}^{\infty}{n^3q^n\over 1-q^n}
\,,\label{E3}\ee
\be
E_6={1\over 2}\left(\vartheta_2^4+\vartheta_3^4\right)
\left(\vartheta_3^4+\vartheta_4^4\right)
\left(\vartheta_4^4-\vartheta_2^4\right)=
1-504\sum_{n=1}^{\infty}{n^5q^n\over 1-q^n}
\,.\label{E4}\ee

The $E_4$ and $E_6$ modular forms have weight four and six,
respectively, and generate the ring of modular forms.  However,
$E_2$ is not exactly a modular form, but \be \hat E_2=E_2-{3\over
  \pi\tau_2}
\label{E9}\ee
is a modular form of weight 2 (though not holomorphic anymore).
The (modular-invariant) $j$ function and $\eta^{24}$ can be written
as
\be
j={E_4^3\over \eta^{24}}={1\over q}+744+\ldots\;\;\;,\;\;\;
\eta^{24}={1\over 2^6\cdot 3^3}\left[E_4^3-E_6^2\right]
\,.\label{E10}\ee

We will also introduce the covariant derivative on modular forms:
\be
F_{d+2}=\left({i\over \pi}\partial_{\tau}+{d/2\over
\pi\tau_2}\right)F_{d}
\equiv D_d\;F_d\;.
\label{E10-bis}\ee
$F_{d+2}$ is a modular form of weight $d+2$ if $F_d$ has weight $d$.
The covariant derivative introduced above has the following
distributive
property:
\be
D_{d_1+d_2}~(F_{d_1}~F_{d_2})=F_{d_{2}}(D_{d_1}F_{d_1})+
F_{d_{1}}(D_{d_2}F_{d_2})
\,.\label{E11}\ee
The following relations and (\ref{E11}) allow the computation of any
covariant derivative
\be \label{cov-deriv}
D_2\;\hat E_2={1\over 6}E_4-{1\over 6}\hat E_2^2
\;\;\;,\;\;\;
D_4\;E_4={2\over 3}E_6-{2\over 3}\hat E_2\;E_4
\;\;\;,\;\;\;
D_{6}\;E_6=E_4^2-\hat E_2\;E_6
\,.\ee

\end{document}